\newif\ifarxiv
\arxivtrue 

\documentclass[aps,pra,twocolumn,showpacs,longbibliography,groupedaddress]{revtex4-1}

\usepackage{amsmath,amsfonts,amssymb}
\usepackage{graphicx}
\usepackage{braket}

\renewcommand{\vec}[1]{\boldsymbol{#1}}

\newif\iffigures
\figurestrue 

\begin{document}

\title{Classical model of delayed-choice quantum eraser}
\author{Brian R. La Cour}
  \email{blacour@arlut.utexas.edu}
  \affiliation{Applied Research Laboratories, 
               The University of Texas at Austin, 
               P.O. Box 8029, 
               Austin, TX 78713-8029}
\author{Thomas W. Yudichak}
  \affiliation{Applied Research Laboratories, 
               The University of Texas at Austin, 
               P.O. Box 8029, 
               Austin, TX 78713-8029}
\date{\today}


\begin{abstract}
Wheeler's delayed-choice experiment was conceived to illustrate the paradoxical nature of wave-particle duality in quantum mechanics.  In the experiment, quantum light can exhibit either wave-like interference patterns or particle-like anti-correlations, depending upon the (possibly delayed) choice of the experimenter.  A variant known as the quantum eraser uses entangled light to recover the lost interference in a seemingly nonlocal and retrocausal manner.  Although it is believed that this behavior is incompatible with classical physics, here we show that the observed quantum phenomena can be reproduced by adopting a simple deterministic detector model and supposing the existence of a random zero-point electromagnetic field.
\end{abstract}




\maketitle


\section{Introduction}

Wave-particle duality is one of the oldest and most perplexing aspects of quantum theory \cite{Bohr1928}.  Although the wave-like nature of light had been well established by the 19th century, experiments of the early 20th century brought about the notion of light as a particle, what we now call a photon \cite{Lenard1902,Einstein1905,Millikan1916}.  Maintaining this notion of light as being composed of discrete particles can, however, be rather paradoxical at times, as numerous real and \textit{gedanken} experiments have shown \cite{Jacques2007,Ionicioiu2011,Roy2012,Peruzzo2012,Kaiser2012,Ionicioiu2014}.

One particular experiment that has captured recent interest and attention is the delayed-choice quantum eraser.  First conceived by Scully and Dr\"{u}hl in 1982 \cite{Scully1982}, the quantum eraser is a variant of Wheeler's delayed choice experiment in which measurements on one of a pair of entangled light beams are used to recover an interference pattern, and hence wave-like behavior, that would otherwise be lost with the introduction of which-way path information in a Mach-Zehnder interferometer \cite{Kim2000,Walborn2002,Gogo2005,Kastner2019,Dong2020}.  For Wheeler, the delayed-choice experiment was an argument for \emph{anti-realism}, the notion that quantum objects, such as photons, do not have definite, intrinsic properties that are independent of the measurement context \cite{Ma2016}.  Some, however, have interpreted the results of delayed-choice quantum eraser experiments as evidence for a form of retrocausality \cite{Aharonov2005}.

More recently, a series of delayed-choice experiments has been performed that rule out a certain class of non-retrocausal hidden-variable models described by Chaves, Lemos, and Pienaar \cite{Chaves2018}.  The general model they describe can provide a causal description of the standard delayed-choice experiment, but it fails to describe a variant of this experiment using variable phase delays in the arms of the interferometer.  This variant has been the subject of recent experimental investigations, which are consistent with theoretical predictions \cite{Polino2019,Huang2019,Pan2019}.  These experiments place certain dimensional restrictions on the class of non-retrocausal hidden variable models that can be consistent with theory and observations.

In this paper, we revisit Wheeler's delayed-choice experiment, its recent experimental variants, and the more elaborate quantum eraser experiment within the context of a simple, physically motivated classical model \cite{LaCour&Williamson2020,LaCour&Yudichak2021}.  Although loophole-free experiments have already been performed to rule out local realism \cite{Hensen2015,Giustina2015,Shalm2015}, the identification of precisely which phenomena defy a classical interpretation remains an open question and one of practical relevance to ensure the security and efficacy of emergement quantum technologies.  Our approach is modeled after stochastic electrodynamics (SED) in assuming a reified zero-point field (ZPF) corresponding to the vacuum state \cite{QuantumDice,EmergingQuantum}.  A significant departure from standard SED in our approach is the introduction of a deterministic model of detectors using an amplitude threshold crossing scheme.  We find that these simple assumptions, combined with standard experimental post-selection and data analysis techniques, adequately describe the observed quantum phenomena.

The structure of the paper is as follows:  In Sec.\ \ref{sec:DC} we consider a simple delayed-choice experiment using weak coherent light as a notional single-photon source.  We replace the coherent light with a source of entangled light in Sec.\ \ref{sec:QE}, within the context of a quantum eraser experiment, and demonstrate how post-selection, not causality, is the mechanism whereby path information is effectively erased.  With these two results established, we revisit the theoretical arguments of Chaves and Bowles in Sec.\ \ref{sec:DW} and argue that their assumptions are overly restrictive.  Section \ref{sec:RE} considers a variant of these experiments using entangled light sources and shows that these, too, can be understood within a classical framework.  Finally, we summarize our conclusions in Sec.\ \ref{sec:fin}.  Figures and numerical experiments were created and performed using a custom simulation tool, the Virtual Quantum Optics Laboratory (VQOL) \cite{VQOL}.


\section{Simple Delayed-Choice Experiment}
\label{sec:DC}

\begin{figure}[h]
\centering
\iffigures
\includegraphics[width=\columnwidth]{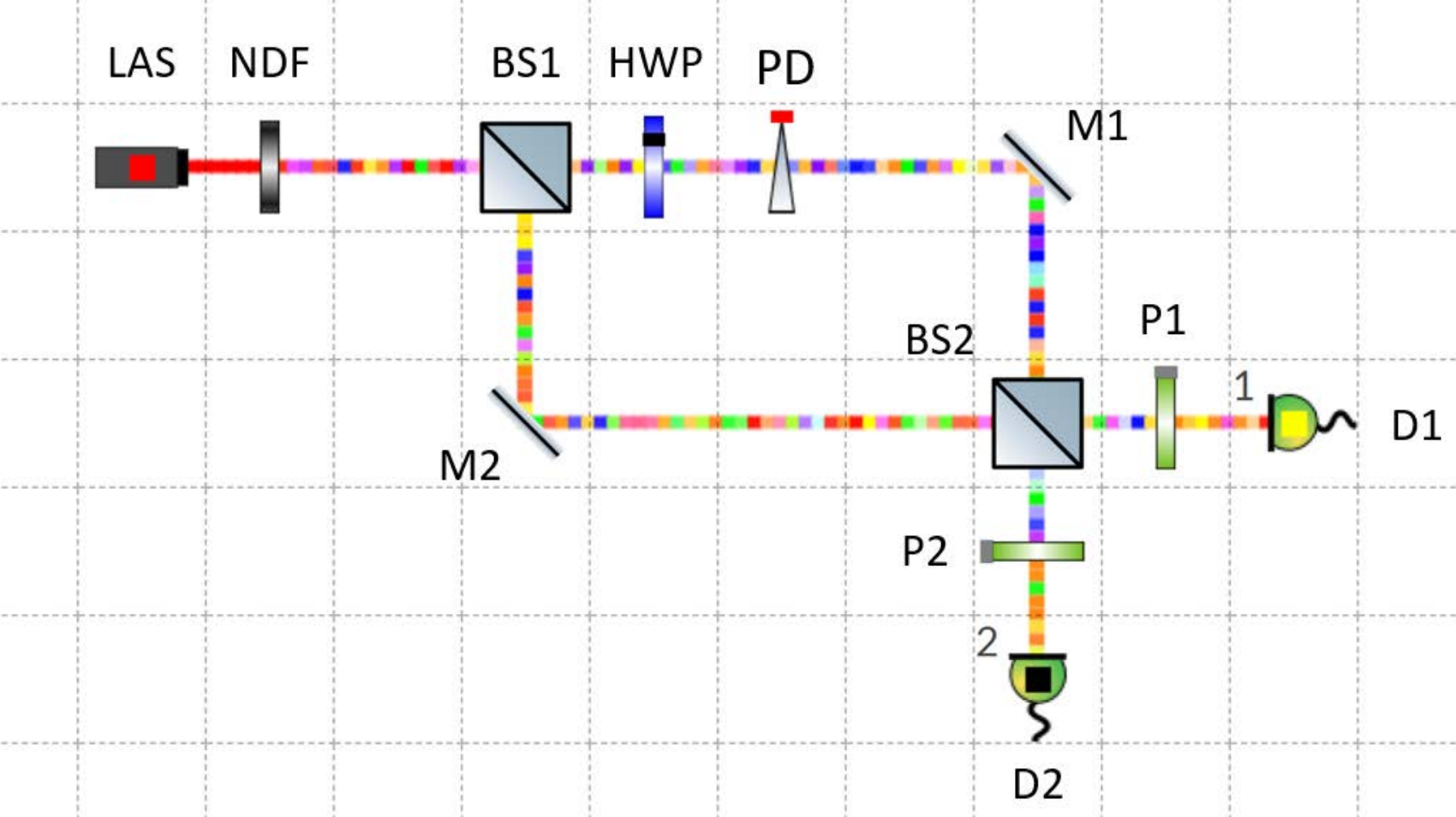}
\fi
\caption{(Color online)  VQOL experimental setup for a simple delayed-choice experiment.  The different colors (or shades of gray) in the beams correspond to different polarizations.}
\label{fig:DCE_setup}
\end{figure}

Consider the Mach-Zehnder interferometer of Fig.\ \ref{fig:DCE_setup}.  A laser (LAS) provides a source of coherent, horizontally polarized light that is strongly attenuated by a neutral density filter (NDF) before entering the first beam splitter (BS1).  Under our model, the laser light exiting the NDF is represented by a stochastic Jones vector of the form
\begin{equation}
\vec{a} = \begin{pmatrix} a_H \\ a_V \end{pmatrix} = \begin{pmatrix} \alpha \\ 0 \end{pmatrix} + \sigma_0 \begin{pmatrix} z_{1H} \\ z_{1V} \end{pmatrix} \; ,
\end{equation}
where $\alpha \in \mathbb{C}$ describes the mean amplitude and phase of the light, $\sigma_0 = 1/\sqrt{2}$ is the scale of the ZPF (corresponding to a modal energy of $\frac{1}{2}\hbar\omega$), and $z_{1H}, z_{1V}$ are independent standard complex Gaussian random variables.  (We say that $z$ is a standard complex Gaussian random variable if $\mathsf{E}[z] = 0$, $\mathsf{E}[|z|^2] = 1$, and $\mathsf{E}[z^2] = 0$.)  Note that $z_{1H}$ and $z_{1V}$ play the role of hidden variables.  This model is mathematically equivalent to the corresponding quantum coherent state $\ket{\alpha} \otimes \ket{0}$ whose Wigner function is a bivariate Gaussian probability density function identical to that of $\vec{a}$.  The effect of the NDF is to ensure that $|\alpha| \ll 1$.

Passage through the first 50/50 beam splitter (denoted by BS1) splits the beam into two orthogonal spatial modes, a right-traveling mode (denoted by $\rightarrow$) and a down-traveling mode (denoted by $\downarrow$).  In addition, there is a down-traveling vacuum mode that enters the top input port of BS1, represented by the independent stochastic Jones vector
\begin{equation}
\vec{b} = \begin{pmatrix} b_H \\ b_V \\ \end{pmatrix} = \sigma_0 \begin{pmatrix} z_{2H} \\ z_{2V} \end{pmatrix} \; ,
\end{equation}
where $z_{2H}, z_{2V}$ are independent standard complex Gaussian random variables (that are also independent of $z_{1H}, z_{1V}$).  The two spatial modes may be represented by a pair of stacked Jones vectors as follows:
\begin{equation}
\begin{bmatrix} a_H \\ a_V \\ -- \\ b_H \\ b_V \end{bmatrix} 
\xrightarrow{\sf BS1}
\frac{1}{\sqrt{2}} \begin{bmatrix} a_H + b_H \\ a_V + b_V \\ ---- \\ a_H - b_H \\ a_V - b_V \end{bmatrix} \; .
\end{equation}
Note that the second term on the right-hand side is again a standard complex Gaussian random vector, owing to the unitarity of the beam splitter transformation.

The right-traveling beam next undergoes a a transformation via a half-wave plate (denoted by HWP) with a fast-axis angle $\theta \in [0,\pi/4]$ relative to the horizontal axis.  It subsequently undergoes a phase delay (denoted by PD) that applies a global phase angle $\phi \in [0,2\pi]$.  Finally, a pair of mirrors (denoted by M1 and M2) swap the two spatial modes.  The resulting stacked Jones vector after these three transformations is now
\begin{equation}
\frac{1}{\sqrt{2}} \begin{bmatrix} a_H + b_H \\ a_V + b_V \\ ---- \\ a_H - b_H \\ a_V - b_V \end{bmatrix}
\xrightarrow{\sf HWP, \; PD, \; M1, \; M2}
\begin{bmatrix} a'_H \\ a'_V \\ -- \\ b'_H \\ b'_V \end{bmatrix} \; ,
\end{equation}
where $a'_H = (a_H - b_H)/\sqrt{2}$, $a'_V = (a_V - b_V)/\sqrt{2}$, and
\begin{subequations}
\begin{align}
b'_H &= \frac{e^{i\phi}}{\sqrt{2}} \Bigl[ \cos2\theta ( a_H + b_H ) + \sin2\theta (a_V + b_V) \Bigr] \; , \\
b'_V &= \frac{e^{i\phi}}{\sqrt{2}} \Bigl[ \sin2\theta ( a_H + b_H ) - \cos2\theta (a_V + b_V) \Bigr] \; .
\end{align}
\label{eqn:woBS2}
\end{subequations}

In the absence of the second beam splitter (denoted by BS2), the Jones vectors $(a'_H, a'_V)^\mathsf{T}$ and $(b'_H, b'_V)^\mathsf{T}$ will determine if a detection is made on the right-traveling mode (by detector D1) or the down-traveling mode (by detector D2).  We adopt an amplitude threshold crossing scheme as a model to determine whether a given detector makes a detection (or ``clicks'') \cite{LaCour&Williamson2020}.  Under this scheme, a detector clicks if the amplitude of either the horizontal or vertical polarization component of the impinging beam falls above a given threshold $\gamma \ge 0$.

Each detector has placed before it a polarizer (denoted by P1 and P2) oriented to admit horizontally polarized light.  A polarizer can be modeled as a polarizing beam splitter for which one of the output ports is ignored \cite{Agarwal}.  Consequently, an independent vacuum mode will be present in the second input port, resulting in the transformation
\begin{equation}
\begin{pmatrix} a'_H \\ a'_V \end{pmatrix} \xrightarrow{\sf P1} \begin{pmatrix} a'_H \\ \sigma_0 z'_{1V} \end{pmatrix} \; ,
\end{equation}
where $z'_{1V}$ is an independent standard complex Gaussian random variable corresponding to the vacuum mode of the notional second input port.  Similarly, passage through P2 will result in the transformation
\begin{equation}
\begin{pmatrix} b'_H \\ b'_V \end{pmatrix} \xrightarrow{\sf P2} \begin{pmatrix} b'_H \\ \sigma_0 z'_{2V} \end{pmatrix} \; ,
\end{equation}
where $z'_{2V}$ is, again, an independent standard complex Gaussian random variable

Thus, detector D1 clicks, according to this model, if the random variables $z_{1H}, z_{1V}, z_{2H}, z_{2V}, z'_{1V}, z'_{2V}$ are such that they fall with the event set
\begin{equation}
D_1 = \Bigl\{ |a'_H| > \gamma \; \mbox{or} \; |\sigma_0 z'_{1V}| > \gamma \Bigr\} \; .
\end{equation}
Similarly, detector D2 clicks under the event set
\begin{equation}
D_2 = \Bigl\{ |b'_H| > \gamma \; \mbox{or} \; |\sigma_0 z'_{2V}| > \gamma \Bigr\} \; .
\end{equation}
Clearly, the probabilities for these events are independent of the phase angle $\phi$, since only the amplitudes of $a'_H$ and $b'_H$ are considered.  Thus, no interference effects would be observed by simply varying $\phi$.

One may consider the alternative counterfactual case in which the second beam splitter, BS2, is present.  In this case, we have a further transformation,
\begin{equation}
\begin{bmatrix} a'_H \\ a'_V \\ -- \\ b'_H \\ b'_V \end{bmatrix}
\xrightarrow{\sf BS2}
\frac{1}{\sqrt{2}} \begin{bmatrix} a'_H + b'_H \\ a'_V + b'_V\\ ---- \\ a'_H - b'_H \\ a'_V - b'_V \end{bmatrix}
=
\begin{bmatrix} a''_H \\ a''_V \\ -- \\ b''_H \\ b''_V \end{bmatrix} \; ,
\label{eqn:wBS2}
\end{equation}
Detector D1 now clicks under the event
\begin{equation}
D'_1 = \Bigl\{ |a''_H| > \gamma \; \mbox{or} \; |\sigma_0 z'_{1V}| > \gamma \Bigr\} \; ,
\end{equation}
while detector D2 clicks under the event
\begin{equation}
D'_2 = \Bigl\{ |b''_H| > \gamma \; \mbox{or} \; |\sigma_0 z'_{2V}| > \gamma \Bigr\} \; .
\end{equation}
Both $a''_H$ and $b''_H$ contain a relative phase term, resulting in a dependence on $\phi$ for the probabilities of the two events.  Thus, the addition or removal of the second beam splitter may causally create or destroy, respectively, an interference pattern, even if this action is taken well after the light has passed through the first beam splitter.

We note this general qualitative behavior is itself unremarkable and may be observed more directly in the intensities of classical light (for which $|\alpha| \gg 1$ or, equivalently, $\sigma_0 \approx 0$), as in this regime we find
\begin{align}
|a'_H|^2 &\approx \frac{1}{2} |\alpha|^2 \; , \\
|a''_H|^2 &\approx \frac{1}{4} |\alpha|^2 \left| 1 + e^{i\phi} \cos2\theta \right|^2 \; .
\end{align}
We further note that non-zero values of $\theta$ can provide which-way information, insofar as they may completely destroy, for $\theta = \pi/2$, or merely reduce, for $0 < \theta < \pi/2$, the intensity inference pattern.  To observe the more subtle particle-like behavior, we must consider the statistical properties of detection events, which we shall now do.

The random variables $a'_H$, $b'_H$ obey a proper complex Gaussian distribution with expectation values of
\begin{align}
\mathsf{E}[a'_H] &= \frac{\alpha}{\sqrt{2}} \; , \\
\mathsf{E}[b'_H] &= \frac{\alpha}{\sqrt{2}} \cos2\theta
\end{align}
and a common variance of $\sigma_0^2$.  They are furthermore independent, owing the unitarity of the transformations involved.  Thus, $|a'_H|$ and $|b'_H|$ each follow a Rician distribution \cite{Rice1945}.  Since, furthermore, $z_{1V}$ and $z_{2V}$ are also independent, the probability of a single click on detector D1 (and \emph{not} D2) is given by
\begin{equation}
p_1(\theta,\phi) = \Pr[D_1 \cap \bar{D}_2] = \Pr[D_1] \left( 1 - \Pr[D_2] \right) \; ,
\end{equation}
where
\begin{align}
\Pr[D_1] &= 1 - F\left( \frac{\alpha}{\sqrt{2}} \right) F(0) \; , \\
\Pr[D_2] &= 1 - F\left( \frac{\alpha}{\sqrt{2}} \cos2\theta \right) F(0) \; ,
\end{align}
and $F(\alpha) = \Pr[|\alpha+z_H| \le \gamma]$ is given by the Marcum Q function as \cite{Marcum1950,LaCour&Williamson2020}
\begin{equation}
F(\alpha) = 1 - Q_1(\sqrt{2}|\alpha|/\sigma_0, \sqrt{2}\gamma/\sigma_0) \; .
\end{equation}
Note that $Q_1(0,\sqrt{2}\gamma/\sigma_0) = e^{-\gamma^2/\sigma_0^2}$.  

A similar analysis may be used when the second beam splitter is in place, albeit using $a''_H$ and $b''_H$ instead.  The probability of a single click on detector D1 is now
\begin{equation}
p'_1(\theta,\phi) = \Pr[D'_1] \left( 1 - \Pr[D'_2] \right) \; ,
\end{equation}
where
\begin{align}
\Pr[D'_1] &= 1 - F\left( \frac{\alpha}{2} (1+e^{i\phi} \cos2\theta) \right) F(0) \; , \\
\Pr[D'_2] &= 1 - F\left( \frac{\alpha}{2} (1-e^{i\phi} \cos2\theta) \right) F(0) \; .
\end{align}

\begin{figure}[h]
\centering
\iffigures
\includegraphics[width=\columnwidth]{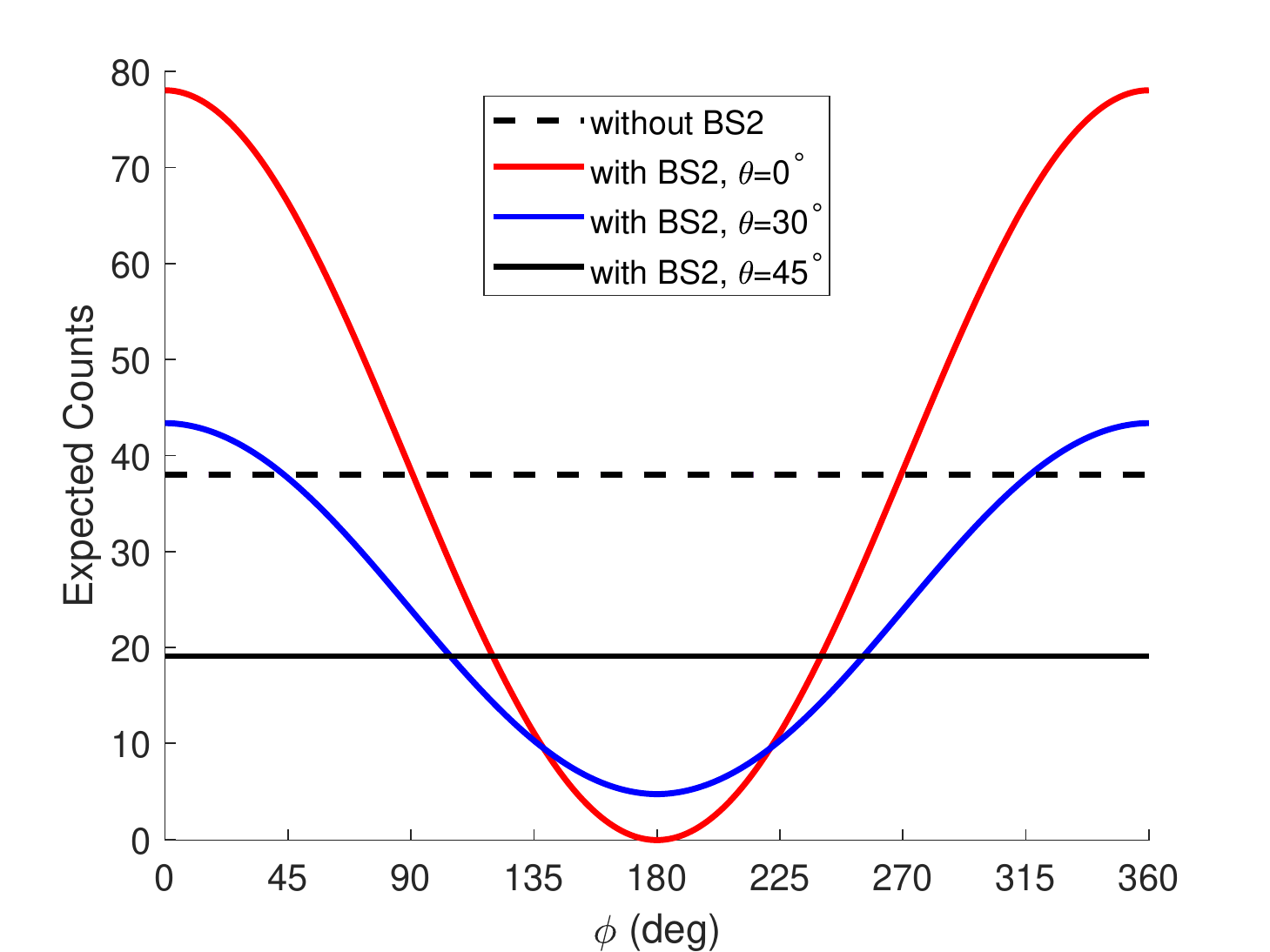}
\fi
\caption{(Color online)  Plot of the expected counts, minus dark counts, for a delayed-choice experiment versus the phase delay angle $\phi$ in the upper arm of the interferometer.  The dashed line corresponds to when BS2 is removed, and the solid lines correspond to when BS2 is present.  Interference patterns are observed when there is no which-way information [$\theta=0^\circ$, red (light gray)] or only partial which-way information [$\theta=30^\circ$, blue (dark gray)].  The interference pattern vanishes with complete which-way information ($\theta=45^\circ$, black).}
\label{fig:DCE_results}
\end{figure}

In Fig.\ \ref{fig:DCE_results} we have plotted the expected number of counts, in excess of the dark counts, for the cases with and without BS2 present and for different values of $\theta$.  The number of notional trials was taken to be $N = 10^6$.  The laser and NDF were taken to be such that $\alpha = 0.1$, and the detectors were such that $\gamma = 1.95$, corresponding to a dark count probability of $p_d = 0.001$.  Without BS2 present, the expected counts are $N p_1(\theta,\phi) - N p_d$; with BS2 in place, they are $N p'_1(\theta,\phi) - N p_d$.  The HWP fast-axis angle, $\theta$, was take to be either 0, 30, or 45 degrees, corresponding to either no, partial, or complete which-way information, respectively.

As expected, there is no interference pattern when BS2 is removed.  With BS2 in place, there is a strong interference pattern when there is no which-way information (i.e., $\theta = 0^\circ$).  When partial which-way information is available, corresponding to a non-zero vertical component of the mean polarization, the interference pattern is diminished but remains discernible.  If complete which-way information is available, corresponding to a mean vertical polarization in the upper arm of the interferometer, the interference pattern is diminished to the point of being no longer present.  These effects are completely causal and arise from the interference of classical, albeit stochastic, waves.  The particle-like behavior, manifested by single-detection events, is a consequence of the low intensity of the input beam, for which detections on both detectors are rare, and the fact that we have removed non-detection events through post-selection.

In the parameter regime we have chosen, the interference pattern matches a scaled and shifted version of the $\cos^2(\phi/2)$ probability predicted by quantum mechanics for a single-photon state.  Larger values of $\alpha$ may exhibit deviations from this prediction, as the corresponding coherent state $\ket{\alpha}$ may no longer be considered a good approximation of a vacuum-plus-single-photon state.

\section{Delayed-Choice Quantum Eraser}
\label{sec:QE}

\begin{figure}[h]
\centering
\iffigures
\includegraphics[width=\columnwidth]{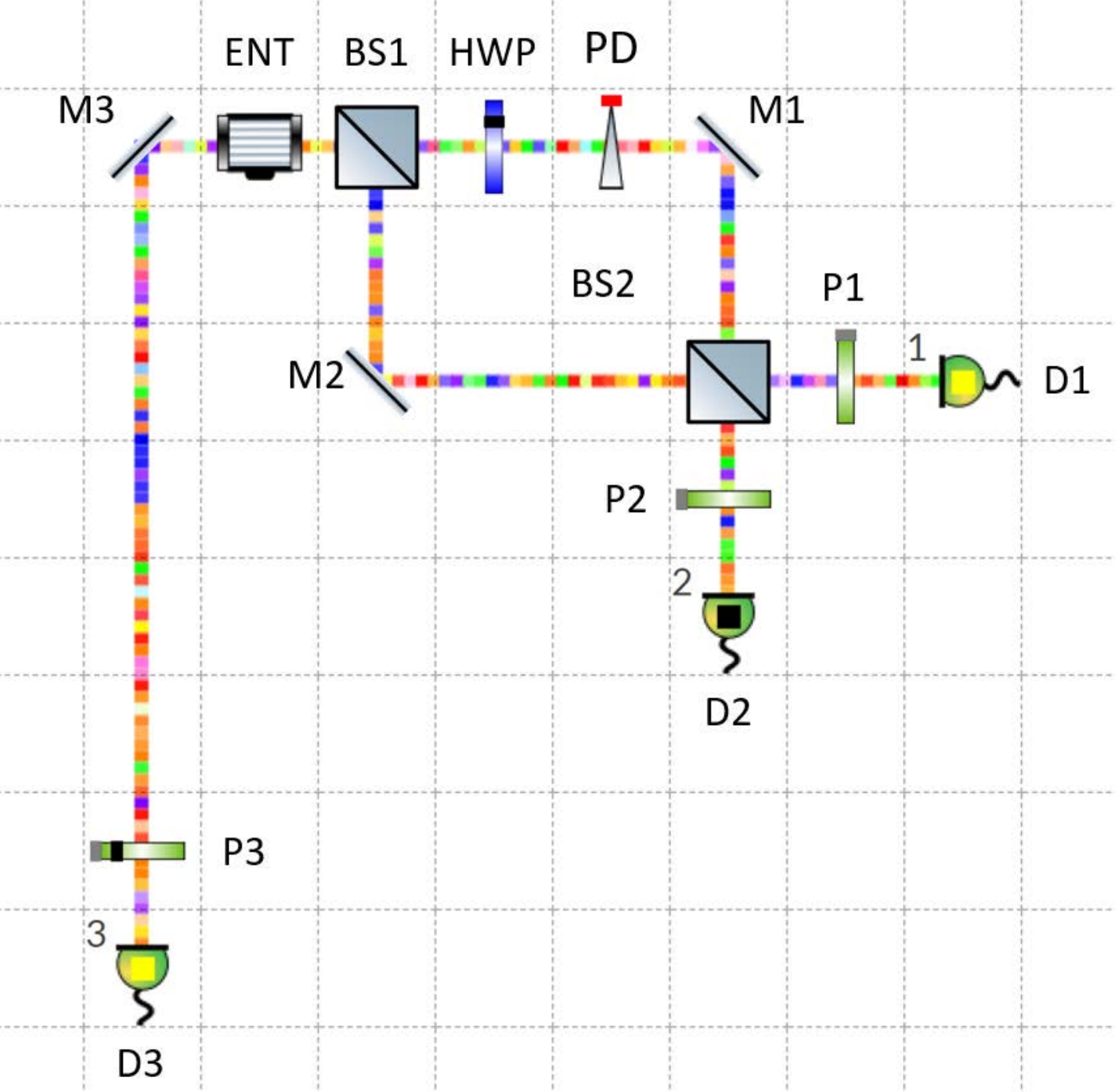}
\fi
\caption{(Color online) VQOL experimental setup for a delayed-choice quantum eraser experiment.  The entanglement source is labeled ENT.}
\label{fig:QE_setup}
\end{figure}

Wheeler's delayed-choice experiment can be changed to a quantum eraser experiment by replacing the laser and NDF with an entanglement source (ENT).  The experiment is illustrated in Fig.\ \ref{fig:QE_setup}, where we have also added a third polarizer and detector, denoted P3 and D3, respectively.  Unlike the other two polarizers, P3 is rotated by $45^\circ$ so as to admit diagonally polarized light.

The entanglement source is modeled classically as a type-I parametric down conversion process for which the inputs states are a pump laser (not shown) and classically modeled vacuum modes from the ZPF \cite{LaCour&Yudichak2021}.  The Jones vectors for the right-traveling ($\rightarrow$) and left-traveling ($\leftarrow$) spatial modes of the entanglement source, denoted $\vec{a}$ and $\vec{c}$, respectively, are given by
\begin{equation}
\vec{a} = \begin{pmatrix} a_H \\ a_V \end{pmatrix}
= \sigma_0 \begin{pmatrix} z_{1H} \cosh r + z_{3H}^* \sinh r \\ z_{1V} \cosh r + z_{3V}^* \sinh r \end{pmatrix} \; \;
\label{eqn:a}
\end{equation}
and
\begin{equation}
\vec{c} = \begin{pmatrix} c_H \\ c_V \end{pmatrix}
= \sigma_0 \begin{pmatrix} z_{3H} \cosh r + z_{1H}^* \sinh r \\ z_{3V} \cosh r + z_{1V}^* \sinh r \end{pmatrix} \; ,
\label{eqn:c}
\end{equation}
where $z_{1H}, z_{1V}, z_{3H}, z_{3V}$ are independent standard complex Gaussian random variables and $r \ge 0$ is the squeezing strength of the entanglement source.  We note that $\vec{a}$ and $\vec{c}$ are statistically dependent for $r > 0$.

The joint probability density function of $\vec{a}$ and $\vec{c}$ is identical to the Gaussian Wigner function of a four-mode entangled  squeezed state \cite{Cahill1969}.  If $r \ll 1$, this squeezed state may be approximated as a superposition of a vacuum state and the entangled Bell state
\begin{equation}
\ket{\Psi} = \frac{\ket{H}\otimes\ket{H} + \ket{V}\otimes\ket{V}}{\sqrt{2}} \; ,
\label{eqn:Bell}
\end{equation}
where the left and right kets in each tensor product indicate the spatial modes traveling to the left and right, respectively.

As before, $\vec{b}$ denotes the Jones vector of the down-traveling ($\downarrow$) vacuum mode entering BS1 and given by
\begin{equation}
\vec{b} = \begin{pmatrix} b_H \\ b_V \end{pmatrix}
= \sigma_0 \begin{pmatrix} z_{2H} \\ z_{2V} \end{pmatrix} \; ,
\label{eqn:b}
\end{equation}
where, again, $z_{2H}, z_{2V}$ are independent standard complex Gaussian random variables.

The transformations of $\vec{a}$ and $\vec{b}$ through the interferometer are identical to those in the simple delayed-choice experiment discussed previously, as given by Eqns.\ (\ref{eqn:woBS2}) and (\ref{eqn:wBS2}).  We shall again denote these by $\vec{a}'', \vec{b}''$ and $\vec{a}', \vec{b}'$ for the cases with and without BS2, respectively.  Passage through the polarizers P1 and P2 results in the transformations
\begin{equation}
\begin{pmatrix} a'_H \\ a'_V \end{pmatrix}
\xrightarrow{\sf P1}
\begin{pmatrix} a'_H \\ \sigma_0 z'_{1V} \end{pmatrix}
\end{equation}
and
\begin{equation}
\begin{pmatrix} b'_H \\ b'_V \end{pmatrix}
\xrightarrow{\sf P2}
\begin{pmatrix} b'_H \\ \sigma_0 z'_{2V} \end{pmatrix} \; ,
\end{equation}
with similar transformations for $\vec{a}'', \vec{b}''$.

Upon passing through polarizer P3, $\vec{c}$ becomes
\begin{equation}
\begin{pmatrix} c_H \\ c_V \end{pmatrix}
\xrightarrow{\sf P3}
\frac{c_D}{\sqrt{2}} \begin{pmatrix} 1 \\ 1 \end{pmatrix} 
+
\frac{\sigma_0 z_{3A}}{\sqrt{2}} \begin{pmatrix} 1 \\ -1 \end{pmatrix} \; ,
\end{equation}
where $c_D = (c_H + c_V)/\sqrt{2}$ is the diagonal component of $\vec{c}$ and $z_{3A}$ is an independent standard complex Gaussian random variable corresponding to the missing anti-diagonal component of the ZPF.

Individual detection events for D1, D2, and D3 for the case in which BS2 is not present may now be defined as follows:
\begin{align}
D_1 &= \Bigl\{ |a'_H| > \gamma \; \mbox{or} \; |\sigma_0 z'_{1V}| > \gamma \Bigr\} \; , \\
D_2 &= \Bigl\{ |b'_H| > \gamma \; \mbox{or} \; |\sigma_0 z'_{2V}| > \gamma \Bigr\} \; , \\
D_3 &= \Bigl\{ \Bigl| \frac{c_D + \sigma_0 z_{3A}}{\sqrt{2}} \Bigr| > \gamma \; \mbox{or} \; \Bigl| \frac{c_D - \sigma_0 z_{3A}}{\sqrt{2}} \Bigr| > \gamma \Bigr\} \; .
\end{align}
The events $D'_1, D'_2, D'_3$, for the case in which BS2 is in place, are defined similarly by replacing $a'_H, b'_H$ with $a''_H, b''_H$.  (Both $c_D$ are $z_{3A}$ are, of course, unchanged.)

Let us consider detection events with BS2 in place and with the HWP rotated to $\theta = 45^\circ$, corresponding to complete which-way information.  Ignoring the detections on detector D3, the probability of a single-detection on detector D1 is, as before,
\begin{equation}
p'_1(\phi) = \Pr[ D'_1 \cap \bar{D}'_2 ] \; .
\end{equation}
If we post-select on events for which detector D3 also clicks, this probability becomes
\begin{equation}
p'_{13}(\phi) = \Pr[ D'_1 \cap \bar{D}'_2 \, | \,  D'_3] \; .
\end{equation}

\begin{figure}[h]
\centering
\iffigures
\includegraphics[width=\columnwidth]{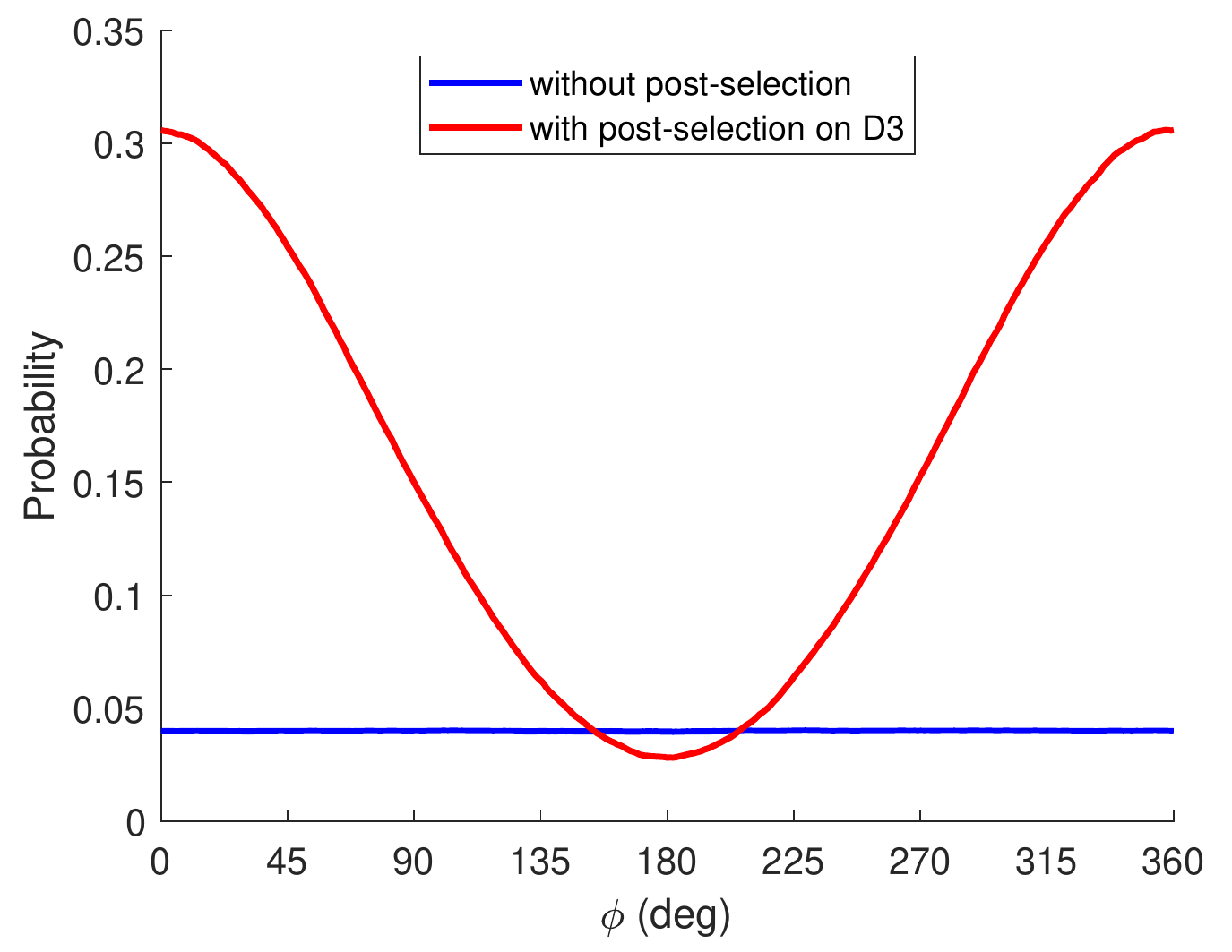}
\fi
\caption{(Color online) Plot of the probability of a single-detection on D1 with either no conditioning on D3 [$p'_1(\phi)$, blue (dark gray) line] or with post-selection of a detection on D3 [$p'_{13}(\phi)$, red (light gray) curve].}
\label{fig:Eraser_results}
\end{figure}

We do not have a closed-form expression for the joint distribution of $\vec{a}$ and $\vec{c}$; however, these probabilities may be estimated numerically.  Taking $\gamma = 1.95$, as before, and $r = 1$, we estimated the probabilities $p'_1(\phi)$ and $p'_{13}(\phi)$ for an ensemble of $N = 10^6$ realizations.  The results are shown in Fig.\ \ref{fig:Eraser_results}.  As expected, the interference pattern is recovered when we post-select on D3 detections.

In the language of quantum mechanics, we have ``erased'' the which-way information by collapsing the state with a projective measurement at D3.  In truth, all we have done is sample from a subensemble of ZPF realizations for which there is a detection on D3.  Classically, there is no causal mechanism by which detection events at D3 affect those at D1 or D2.  It is merely a reflection of the pre-existing correlations between $\vec{a}$ and $\vec{c}$ and their modification as a result of post-selection.  Of course, it does not matter whether BS2 was inserted before or after the light has passed through BS1, so there is no need for retrocausal explanations either.

\section{Quantum-Controlled Experiments}
\label{sec:QDC}

In the quantum eraser experiment of the previous section, the second beam splitter was either present or absent, resulting in the presence or absence of an interference pattern, respectively.  A similar experiment was performed by Jacques \textit{et al.}\ using a quantum random-number generator (QRNG) and a classical switch \cite{Jacques2007}.  Ionicioiu and Terno have suggested using a controlled quantum gate in place of a classical switch and, furthermore, have argued that this scheme may be used to rule out a certain class of hidden-variable models \cite{Ionicioiu2011,Ionicioiu2014}.  Subsequently, this proposal was implemented experimentally using a programmable nuclear magnetic resonance (NMR) device \cite{Roy2012}, a reconfigurable integrated photonic device \cite{Peruzzo2012}, and a polarization-dependent beam splitter (PDBS) \cite{Kaiser2012}.  All experiments showed the expected continuum of wave-like and particle-like behavior for different experimental settings controlling which-way path information.  In this section, we consider a classical model for the PDBS experiment.

\begin{figure}[h]
\centering
\iffigures
\includegraphics[width=\columnwidth]{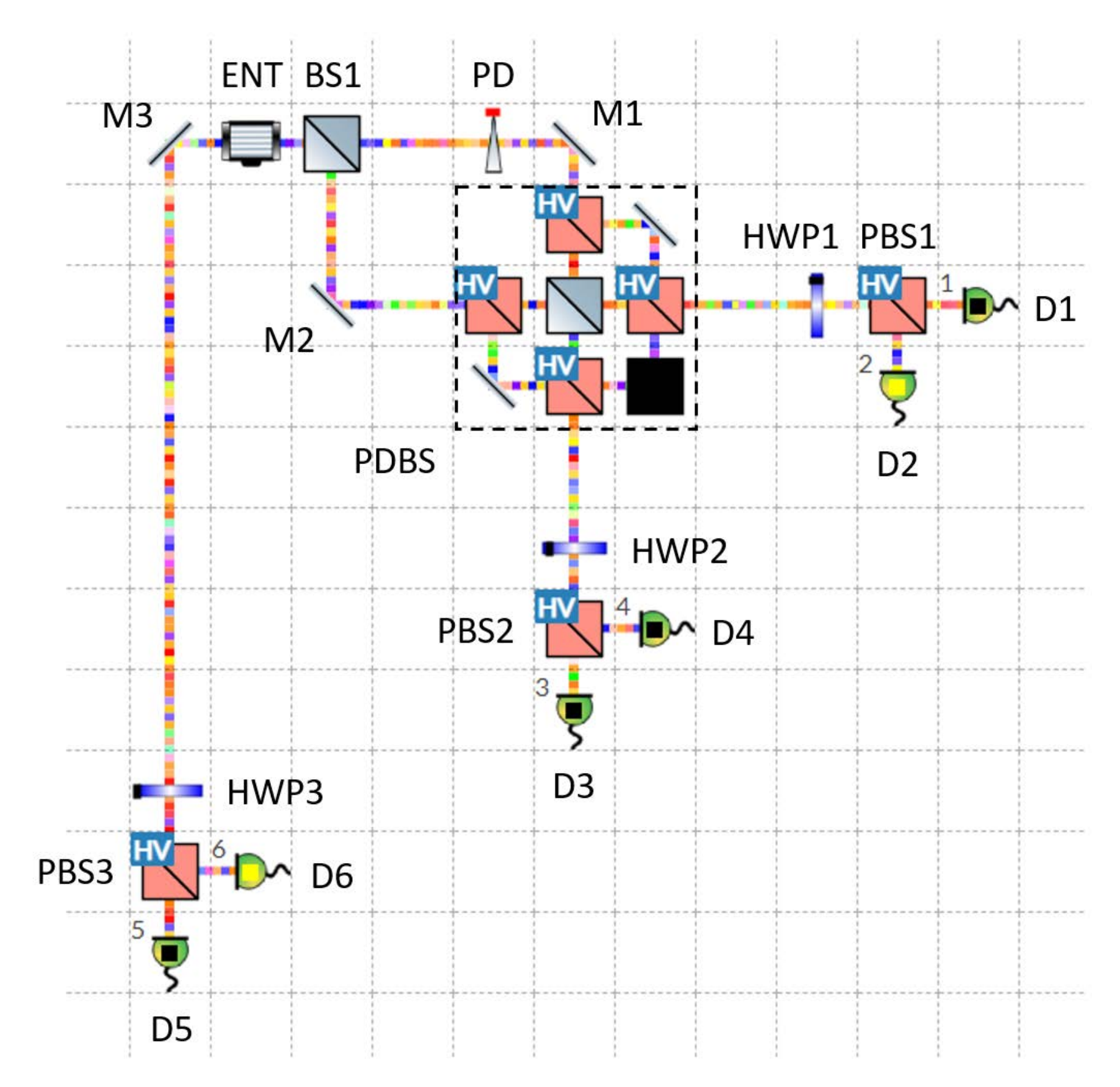}
\fi
\caption{(Color online) VQOL experimental setup for a quantum-controlled delayed-choice quantum eraser experiment using a polarization-dependent beam splitter (PDBS).  The PDBS is composed of a beam splitter surrounded by four polarizing beam splitters that transmit (reflect) horizontal (vertical) light, and a beam blocker (black square).  The beam blocker absorbs ZPF modes entering from the unused input ports of the top and left PBS of the PDBS.}
\label{fig:Kaiser_setup}
\end{figure}

To this end we have implemented the experimental setup described in Ref.\ \cite{Kaiser2012} with a PDBS that behaves as a 50/50 beam splitter for horizontally polarized light and is transparent (except for a swapping of spatial modes) for vertically polarized light.  (See Fig.\ \ref{fig:Kaiser_setup}.)  For Jones vectors $\vec{a}$ and $\vec{b}$ corresponding to light entering from the top and left, respectively, of the PDBS, the transformed polarization states are given by
\begin{equation}
\begin{bmatrix} a'_H \\ a'_V \\ b'_H \\ b'_V \end{bmatrix} = 
\begin{pmatrix}
\frac{1}{\sqrt{2}} & 0 & \;\:\frac{1}{\sqrt{2}} & 0 \\ 0 & 0 & \;\:0 & 1 \\ \frac{1}{\sqrt{2}} & 0 & -\frac{1}{\sqrt{2}} & 0 \\ 1 & 0 & \;\:0 & 0
\end{pmatrix}
\begin{bmatrix} a_H \\ a_V \\ b_H \\ b_V \end{bmatrix} \; .
\end{equation}

As before, the phase delay (PD) is set to an angle of $\phi \in [0,2\pi]$, and we have added a half-wave plate, HWP3, with a fast-axis angle of $\theta \in [0,\pi/4]$.  The other two half-wave plates, HWP1 and HWP2, have fast-axis angles of $\pi/8$ in order to measure the output of the interferometer in the diagonal/anti-diagonal (D/A) basis.  

In quantum mechanical terms, the initial entangled state is given by Eqn.\ (\ref{eqn:Bell}), and the final quantum state, prior to polarizing beam splitters PBS1, PBS2, and PBS3, is given by
\begin{equation}
\begin{split}
\ket{\Psi'} \,=\, &\frac{1}{\sqrt{2}} \left[ \hat{A}_w^\dagger \cos2\theta + \hat{A}_p^\dagger \sin2\theta \right] \hat{c}_H^\dagger \ket{\vec{0}} \\
+  &\frac{1}{\sqrt{2}} \left[ \hat{A}_w^\dagger \sin2\theta - \hat{A}_p^\dagger \cos2\theta \right] \hat{c}_V^\dagger \ket{\vec{0}} \; ,
\end{split}
\end{equation}
where $\ket{\vec{0}}$ is the vacuum state, $\hat{A}_w^\dagger$ is the ``wave'' creation operator
\begin{equation}
\hat{A}_w^\dagger = \frac{1}{2\sqrt{2}} \Bigl[ (1+e^{i\phi}) (\hat{a}_H^\dagger + \hat{a}_V^\dagger) + (1-e^{i\phi}) (\hat{b}_H^\dagger + \hat{b}_V^\dagger) \Bigr] \; ,
\end{equation}
$\hat{A}_p^\dagger$ is the ``particle'' creation operator
\begin{equation}
\hat{A}_p^\dagger = \frac{1}{2} \Bigl[ e^{i\phi} \hat{a}_H^\dagger - e^{i\phi} \hat{a}_V^\dagger + \hat{b}_H^\dagger - \hat{b}_V^\dagger \Bigr] \; ,
\end{equation}
and the operators $\hat{a}_H^\dagger$, $\hat{a}_V^\dagger$, $\hat{b}_H^\dagger$, $\hat{b}_V^\dagger$, $\hat{c}_H^\dagger$, $\hat{c}_V^\dagger$ are the creation operators for the modes corresponding to detectors D1, D2, D3, D4, D5 D6, respectively.

In the quantum mechanical description, a single detection on either D5 or D6 results in a collapse of the wave function onto one of two subspaces, each of which is a superposition of wave-like and particle-like states.  A single detection on D1 or D2, say, conditioned on a single detection on D6 would therefore have a probability of
\begin{equation}
q(\theta,\phi) = \cos^2(\phi/2) \sin^2(2\theta) + \tfrac{1}{2} \cos^2(2\theta) \; .
\label{eqn:Queen}
\end{equation}
Thus, for $\theta = 0$ ($\theta = \pi/4$) we expect fully particle-like (wave-like) behavior.  This result is equivalent to that found in Ref.\ \cite{Kaiser2012}, which uses $\theta$ and $\alpha$ in place of $\phi$ and $2\theta$ and conditions on the horizontal, rather than vertical, mode at PBS3 due to opposite PDBS conventions.

For our classical model we take our initial states to be the Jones vectors $\vec{a}$, $\vec{b}$, and $\vec{c}$ as defined in  Eqns.\ (\ref{eqn:a}), (\ref{eqn:b}), and (\ref{eqn:c}), respectively.  The final Jones vectors for the light entering PBS1, PBS2, and PBS3 are then found to be $\vec{a}'$, $\vec{b}'$, and $\vec{c}'$, respectively, where $\vec{a}'$ is given by
\begin{subequations}
\begin{align}
a'_H &= \frac{1+e^{i\phi}}{2\sqrt{2}} a_H + \frac{e^{i\phi}}{2} a_V - \frac{1-e^{i\phi}}{2\sqrt{2}} b_H + \frac{e^{i\phi}}{2} b_V \; , \\
a'_V &= \frac{1+e^{i\phi}}{2\sqrt{2}} a_H - \frac{e^{i\phi}}{2} a_V - \frac{1-e^{i\phi}}{2\sqrt{2}} b_H - \frac{e^{i\phi}}{2} b_V \; , 
\end{align}
\end{subequations}
$\vec{b}'$ is given by
\begin{subequations}
\begin{align}
b'_H &= \frac{1-e^{i\phi}}{2\sqrt{2}} a_H + \frac{1}{2} a_V - \frac{1+e^{i\phi}}{2\sqrt{2}} b_H - \frac{1}{2} b_V \; , \\
b'_V &= \frac{1-e^{i\phi}}{2\sqrt{2}} a_H - \frac{1}{2} a_V - \frac{1+e^{i\phi}}{2\sqrt{2}} b_H + \frac{1}{2} b_V \; , 
\end{align}
\end{subequations}
and $\vec{c}'$ is given by
\begin{subequations}
\begin{align}
c'_H &= c_H \cos2\theta + c_V \sin2\theta \; , \\
c'_V &= c_H \sin2\theta - c_V \cos2\theta \; .
\end{align}
\end{subequations}

Unlike the quantum description, there is no clear distinction between modal subspaces.  For $\theta = 0$, say, a single detection on D6 will occur when $|c_V|$ is large and $|c_H|$ is small.  Since $\vec{c}$ and $\vec{a}$ are statistically correlated, this will tend to occur when $|a_V|$ is large and $|a_H|$ is small as well.  This, in turn, implies that $a'_H$ and $a'_V$ are dominated by the particle-like $a_V$ term, vice the wave-like $a_H$ term, thereby giving rise to more particle-like behavior.  However, the effects of the wave-like $a_H$ term will not be completely absent, particularly since, in amplitude, it may be up to a factor of $\sqrt{2}$ larger than the particle-like $a_V$ term.  Thus, some wave-like interference will persist, and a mixture of wave-like and particle-like behavior will always be exhibited.

As before, we may define the individual detection events for each detector as follows:
\begin{subequations}
\begin{align}
D_1 &= \Bigl\{ |a'_H| > \gamma \; \mbox{or} \; |\sigma_0 z'_{1V}| > \gamma \Bigr\} \\
D_2 &= \Bigl\{ |a'_V| > \gamma \; \mbox{or} \; |\sigma_0 z'_{1H}| > \gamma \Bigr\} \\
&\vdots \nonumber \\
\setcounter{equation}{5}
D_6 &= \Bigl\{ |c'_V| > \gamma \; \mbox{or} \; |\sigma_0 z'_{3H}| > \gamma \Bigr\} \; ,
\end{align}
\end{subequations}
where $z'_{1H}, \ldots, z'_{3V}$ are independent standard complex Gaussian random variables corresponding to the unused input ports of PBS1, PBS2, and PBS3.

We may now define the four relevant coincident detection events, $C_{16}$, $C_{26}$, $C_{36}$, and $C_{46}$, as follows:
\begin{subequations}
\begin{align}
C_{16} &= D_1 \cap \bar{D}_2 \cap \bar{D}_3 \cap \bar{D}_4 \cap \bar{D}_5 \cap D_6 \\
&\vdots \nonumber \\
\setcounter{equation}{3}
C_{46} &= \bar{D}_1 \cap \bar{D}_2 \cap \bar{D}_3 \cap D_4 \cap \bar{D}_5 \cap D_6
\end{align}
\end{subequations}
Finally, the conditional probability of a single detection on D1 or D2, given a coincident detection with D6, is 
\begin{equation}
p(\theta,\phi) = \frac{\Pr[C_{16} \cup C_{26}]}{\Pr[C_{16} \cup C_{26} \cup C_{36} \cup C_{46}]} \; .
\end{equation}
Our task is now to compare $p(\theta,\phi)$ to $q(\theta,\phi)$.

\begin{figure}[h]
\centering
\iffigures
\includegraphics[width=\columnwidth]{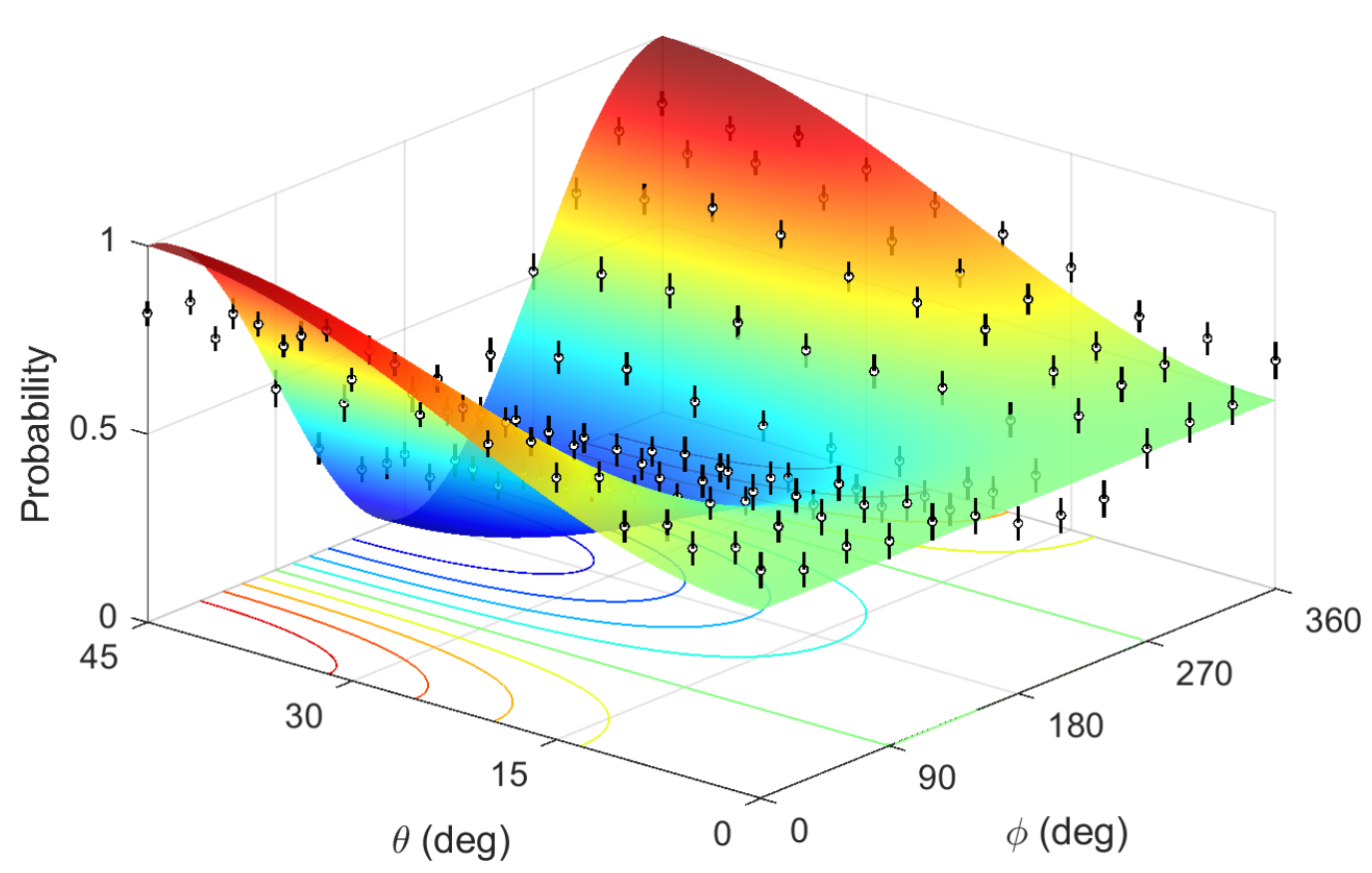}
\fi
\caption{(Color online) Surface plot of $q(\theta,\phi)$ and point estimates of $p(\theta,\phi)$ (black circles) based on numerical experiments.  The small vertical lines represent approximate 95\% confidence intervals.  No fitting was performed on the data.}
\label{fig:morph}
\end{figure}

To do this, we generated $N = 5 \times 10^6$ random realizations for each $\theta$ and $\phi$, corresponding 5 s of simulation time in VQOL.  We took $\gamma = 1.95$ and $r = 0.25$, to better match the results of Ref.\ \cite{Kaiser2012}.  For these settings we found an average of 125 coincidences per run, about five of which may be considered accidental.  In Fig.\ \ref{fig:morph} we have plotted $q(\theta,\phi)$ as well as estimates of $p(\theta,\phi)$ from our numerical experiments.  We find that the results compare favorably with Fig.\ 4 of Ref.\ \cite{Kaiser2012} and exhibit a clear morphing between particle-like and wave-like behavior, although there are deviations from the ideal quantum prediction.  We note also the slight vestige of wave-like behavior at $\theta = 0$ and particle-like behavior at $\theta = \pi/4$, indicating that these results are qualitatively similar to, albeit quantitatively different from, the ideal two-photon quantum predictions.

Given these results, we should like to return to the proof by Ionicioiu and Terno that conformity with the quantum predictions should rule out a certain class of hidden-variable models, as this would seem to be at variance with our results.  This class is specified by three defining characteristics: (1) wave-particle objectivity, (2) determinism, and (3) local independence \cite{Ionicioiu2011,Ionicioiu2014}.

Our model is clearly deterministic: given a particular realization of all ZPF modes, the detection outcomes are uniquely determined.  It is approximately consistent with wave-particle objectivity, which states that the set of hidden-variable states corresponding to the presence or absence of a second beam splitter should be disjoint.  The notional idea is that there is a hidden variable that determines whether particle-like or wave-like behavior is exhibited.  Certainly the sets of hidden-variable states for which a detection occurs on either D5 or D6 (i.e., the sets $D_5 \cap \bar{D}_6$ and $\bar{D}_5 \cap D_6$) are disjoint, but conditioning on these events provides only approximate conformity with the ideal two-photon quantum predictions.  Even under conditioning, the outcomes are a mixture of particle-like and wave-like behavior.

Finally, local independence is the property that the hidden-variable space may be separated into two sets of statistically independent variables controlling, on the one hand, detections at the output of the interferometer and, on the other, the presence or absence of the second beam splitter.  Since the experimental setup is not a proper implementation of the Ionicioiu-Terno scheme, this condition cannot be satisfied.  Specifically, the right spatial mode of the entanglement source is used both as a control and an input to the interferometer.  Although the ZPF modes, which may be construed as the hidden variables, are all statistically independent, they cannot be uniquely associated with either control or input due to the experimental setup and, in particular, use of the PDBS.  Of course, the presence or absence of the second beam splitter could be controlled by a separate, independent classical random variable, but this reduces to the quantum eraser experiment of the previous section.


\section{Relation to Dimension Witness}
\label{sec:DW}

We now turn to some more recent theoretical and experimental results regarding delayed-choice experiments.  Bowles, Quintino, and Brunner have considered the problem of distinguishing quantum systems from classical counterfeits in a device-independent manner using a so-called ``dimension witness'' \cite{Bowles2014}.  In their scheme, the classical system is assumed to produce, upon a certain state preparation, a ``message'' $m \in \{0, \ldots, d-1\}$, where $d$ is the Hilbert space dimension of the corresponding quantum system.  (``Classical,'' in their sense, means that only finite, digital information is conveyed.)  The final measurement outcome is assumed to depend only on the message $m$ and some given measurement setting.

Denoting by $p(x,y)$ the probability of a desired outcome for preparation $x$ and measurement $y$, both taken to be integer indexes, Bowles \emph{et al.}\ define the matrix
\begin{equation}
W_2 = \begin{pmatrix}
p(1,1) - p(2,1) & p(3,1) - p(4,1) \\
p(1,2) - p(2,2) & p(3,2) - p(4,2)
\end{pmatrix} \; .
\end{equation}
For a two-dimensional classical system (i.e., one for which messages are restricted to one of $d = 2$ values), they show that only $\det(W_2) =  0$ is possible.  By contrast, a two-dimensional quantum system (i.e., one for which the Hilbert space dimension is $d = 2$) can achieve values as high as $|\det(W_2)| = 1$.

Chaves \emph{et al.}\ have used this construction to design a delayed-choice experiment that can distinguish between a two-dimensional quantum system and a classical system, so defined, of the same dimension \cite{Chaves2018}.  One version of the proposed experiment was recently performed using a single qubit in a polarization-based Mach-Zehnder interferometer and a pair of phase retarders to provide different preparation and measurement settings \cite{Polino2019}.

\begin{figure}[h]
\centering
\iffigures
\includegraphics[width=\columnwidth]{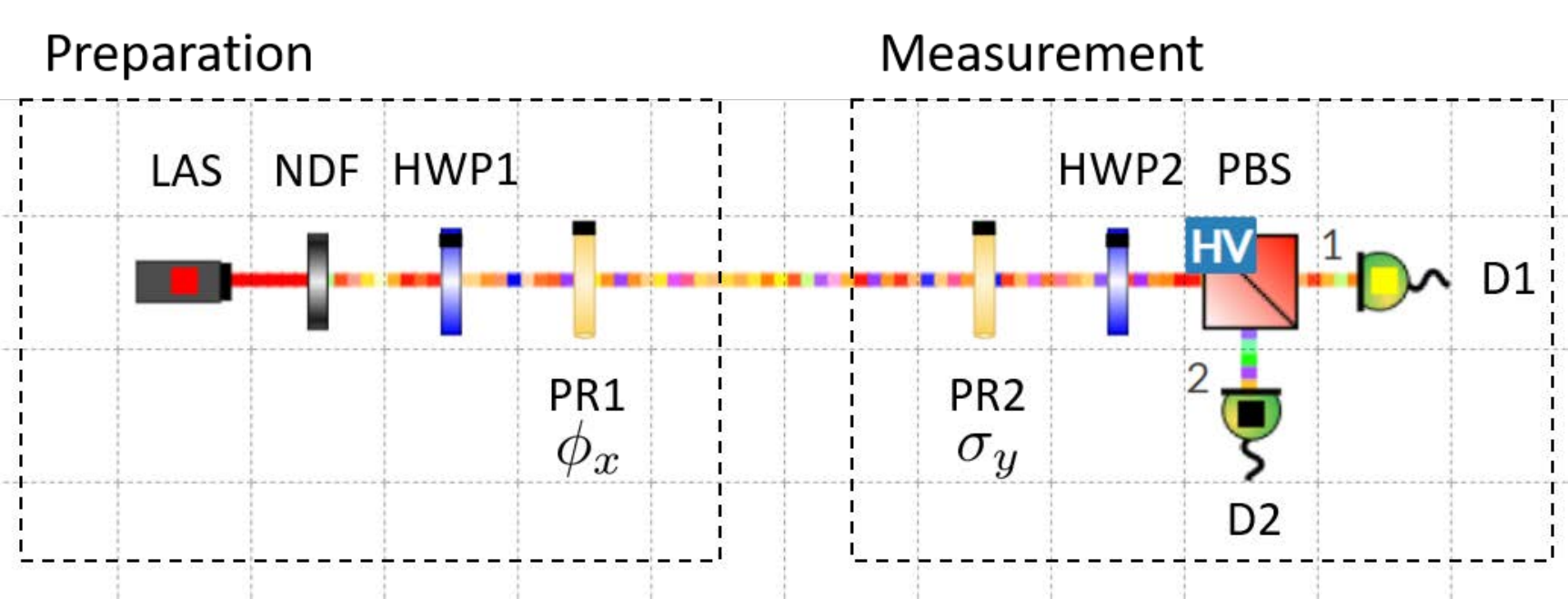}
\fi
\caption{(Color online) VQOL experimental setup for dimension witness experiment.}
\label{fig:DW_setup}
\end{figure}

A simplified version of the experimental setup is shown in Fig.\ \ref{fig:DW_setup}.  A laser (LAS) generates horizontally polarized light, which is attenuated by a neutral density filter (NDF).  A half-wave plate (HWP1) with a fast-axis angle of $22.5^\circ$ is then applied to play the role of a beam splitter.  Next, a birefringent phase retarder (PR1) applies a phase factor $e^{i\phi_x}$ to the vertical polarization component.  This constitutes the preparation stage of the experiment.  The measurement stage consists of another phase retarder, PR2, that applies a phase factor $e^{i\sigma_y}$ to the vertical component.  A second half-wave plate, HWP2, rotated $22.5^\circ$ completes the interferometer, and a final polarizing beam splitter (PBS) and two detectors, D1 and D2, are used to detect horizontal and vertical polarization, respectively.  In the actual experiment, $\sigma_y$ was chosen randomly and set only after the light had passed through HWP1.

In our model, the initial state following the NDF is represented, as before, by the stochastic Jones vector
\begin{equation}
\begin{pmatrix} a_H \\ a_V \end{pmatrix} = \begin{pmatrix} \alpha \\ 0 \end{pmatrix} + \sigma_0 \begin{pmatrix} z_{1H} \\ z_{1V} \end{pmatrix} \; .
\end{equation}
After passing through the phase retarders and HWP2, the state becomes
\begin{equation}
\begin{pmatrix} a_H \\ a_V \end{pmatrix}
\xrightarrow{\sf MZI}
\frac{1}{2} \begin{bmatrix} (1+e^{i\Delta_{xy}}) a_H + (1-e^{i\Delta_{xy}}) a_V \\ (1-e^{i\Delta_{xy}}) a_H + (1+e^{i\Delta_{xy}}) a_V \end{bmatrix} \; ,
\end{equation}
where $\Delta_{xy} = \phi_x + \sigma_y$.

For each preparation $x$ and measurement $y$, we define $p_1(x,y)$ as the probability of obtaining a detection on D1, given that a single detection occurred on either D1 or D2.  This probability is given by
\begin{equation}
p_1(x,y) = \frac{\Pr[D_1\cap\bar{D}_2]}{\Pr[D_1\cap\bar{D}_2] + \Pr[\bar{D}_1\cap D_2]} \; ,
\end{equation}
where
\begin{align}
D_1 &= D_{+} \cup \{ |\sigma_0 z_{2V}| > \gamma \} \\
D_2 &= D_{-} \cup \{ |\sigma_0 z_{2H}| > \gamma \}
\end{align}
and
\begin{equation}
D_{\pm} = \left\{ \left| (1 \pm e^{i\Delta_{xy}}) a_H + (1 \mp e^{i\Delta_{xy}}) a_V \right| > \gamma \right\} \; .
\end{equation}
We similarly define $p_2(x,y)$ as the probability of a single detection on D2.  Note that $z_{2H}$ and $z_{2V}$ correspond to the ZPF components entering from the top input port of the PBS and constitute an independent source of randomness in the measurement stage.

Following the experimenters, we construct $W_2$ for the four preparation settings $\phi_1 = 0$, $\phi_2 = \pi$, $\phi_3 = -\pi/2$, $\phi_4 = \pi/2$ and two measurement settings $\sigma_1 = 0$, $\sigma_2 = \pi/2$.  Quantum mechanics predicts, for a single photon, that $p(1,1) = p(3,2) = 1$, $p(2,1) = p(4,2) = 0$, and $p(3,1) = p(4,1) = p(1,2) = p(2,2) = \frac{1}{2}$, giving $|\det(W_2)| = 1$.  Using our model, we take $p(x,y) = p_1(x,y)$ and compute the matrix $W_2$ using $\sigma_0 = 1/\sqrt{2}$ and $\gamma = 1.95$, as before, and vary $\alpha$.

\begin{figure}[h]
\centering
\iffigures
\includegraphics[width=\columnwidth]{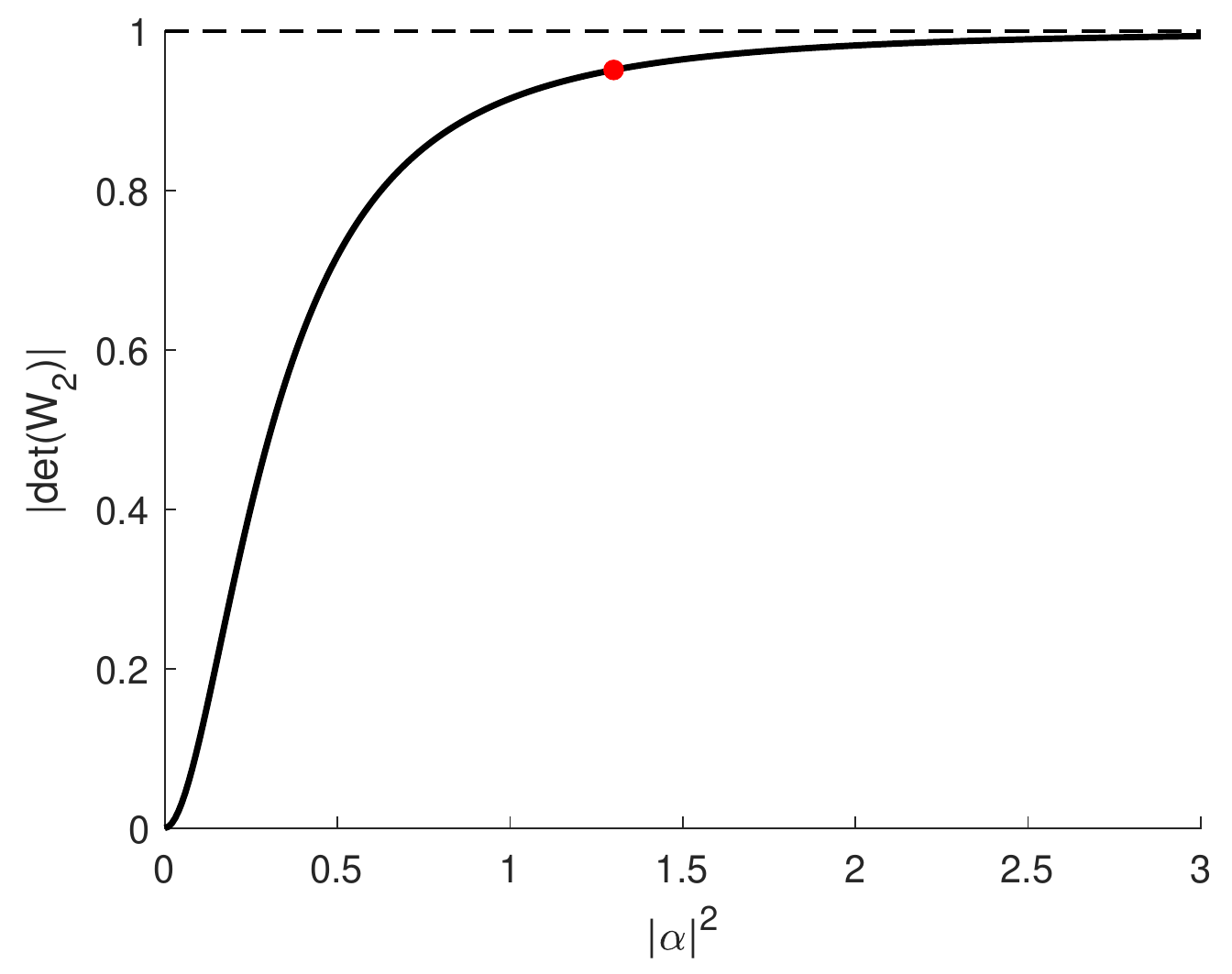}
\fi
\caption{(Color online) Plot of $|\det(W_2)|$ versus $|\alpha|^2$.  The dashed line indicates the ideal quantum prediction.  The red (light gray) dot corresponds to the value of $|\det(W_2)|$ obtained experimentally in Ref.\ \cite{Polino2019}.}
\label{fig:DW_results}
\end{figure}

In Fig.\ \ref{fig:DW_results} we have plotted our classical model results for $|\det(W_2)|$ as a function of the input intensity, as given by $|\alpha|^2$.  (For a coherent state $\ket{\alpha}$, this corresponds to the average photon number.)  For  $0 < |\alpha|^2 \ll 1$, the ZPF dominates and the low coherence results in a small, but positive, value of $|\det(W_2)|$.  For $|\alpha|^2 \gg 1$, $|\det(W_2)|$ approaches the ideal single-photon prediction of one.  Note that, although the limit $|\alpha|^2 \to \infty$ corresponds to classical light (i.e., light for which the ZPF is negligible), post-selection on single-detection events maintains the particle-like anti-correlations of the interferometer and allows for agreement with the quantum prediction.  The experimental value of 0.95 observed in Ref.\ \cite{Polino2019} corresponds, in our model, to $|\alpha|^2 = 1.3$.

Of course, it is unsurprising that a value of $|\det(W_2)|$ greater than zero can be obtained, as the ``messages'' between preparation and measurement in our model are not limited to discrete values but, rather, can take on a continuum of possible values in $\mathbb{C}^2$.  Agreement with quantum mechanics was further made possible by post-selecting on what would appear to be single-photon detections, as was done in the actual experiment.  It is clear from this example that an assumption of finite-dimensional classical models is overly restrictive.

In our model, the measurement outcomes depend on what may be construed as four hidden variables: $z_{1H}$, $z_{1V}$, $z_{2H}$, and $z_{2V}$.  The first two, $\lambda = (z_{1H}, z_{1V})$, may be considered as part of the preparation stage, while the last two, $(z_{2H},z_{2V})$ may be considered part of the measurement stage.  They are independent of and unaffected by the choice of preparation or measurement settings, so it is clear that the model is completely causal.  Nevertheless, it may be useful to examine measurable bounds on retrocausality.

To study this question, Chaves \emph{et al.}\ have proposed the use of a retrocausality measure $R_{Y \to \Lambda}$ intended to capture the apparent retrocausal impact of the measurement setting choice on the preparation hidden-variable state \cite{Chaves2018}.  A hidden-variable model for which $R_{Y \to \Lambda} = 0$ is completely causal, while one for which $R_{Y \to \Lambda} = 1$ is considered strongly retrocausal.  Since, in our model, the preparation hidden variables are independent of the measurement settings, $R_{Y \to \Lambda} = 0$.  Chaves \emph{et al.}\ show that one can measure a bound $R_{\rm min}$ such that any hidden-variable model must satisfy $R_{Y \to \Lambda} \ge R_{\rm min}$ in order to be consistent with observations.  Thus, they conclude, if $R_{\min} > 0$ then such a model must be retrocausal.

The bound $R_{\rm min}$ is given in terms of a dimension witness $I_{\rm DW}$ \cite{Gallego2010}.  This, in turn, is defined to be
\begin{equation}
I_{\rm DW} = \bigl| \braket{B}_{11} + \braket{B}_{12} + \braket{B}_{21} - \braket{B}_{22} - \braket{B}_{31} \bigr| \; ,
\end{equation}
where $\braket{B}_{xy} = p_1(x,y) - p_2(x,y)$.  In terms of $I_{\rm DW}$,
\begin{equation}
R_{\rm min} = \max\left\{ \frac{I_{\rm DW}-3}{4}, \; 0 \right\} \; .
\end{equation}

Algebraically, $I_{\rm DW} \in [0, 5]$ and $R_{\rm min} \in [0, \frac{1}{2}]$.  It has been shown that $I_{\rm DW} \le 3$ for two-dimensional classical systems, while a two-dimensional quantum system may achieve up to $I_{\rm DW} = 1+2\sqrt{2} \approx 3.8284$ \cite{Ahrens2012}.  Consequently, $R_{\rm min} = 0$ for two-dimensional classical systems, while a two-dimensional quantum system may achieve up to $R_{\rm min} = (\sqrt{2}-1)/2 \approx 0.2071$.

\begin{figure}[h]
\centering
\iffigures
\includegraphics[width=\columnwidth]{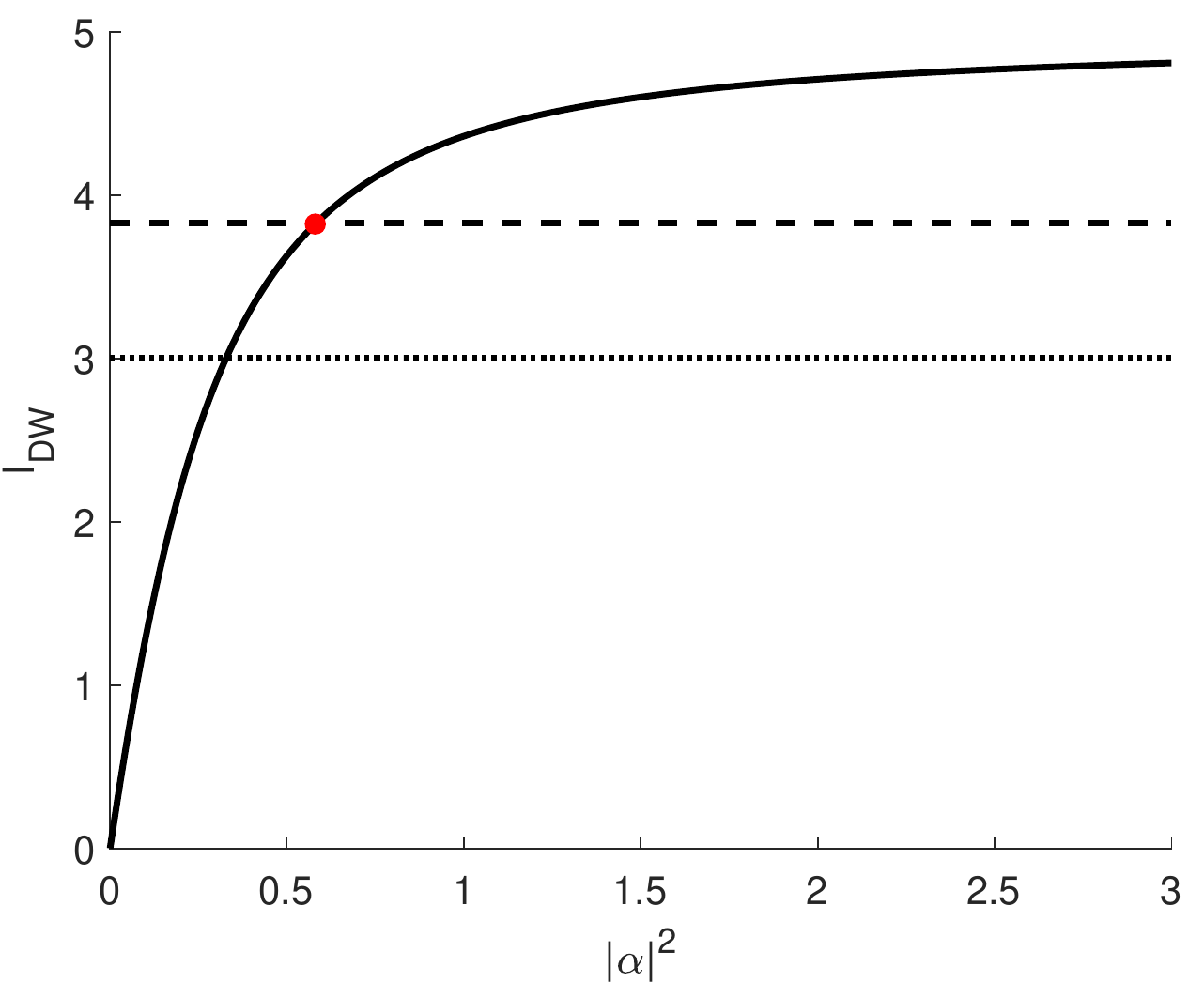}
\fi
\caption{(Color online) Plot of the dimension witness $I_{\rm DW}$ versus $|\alpha|^2$.  The upper dashed line indicates the ideal, single-photon quantum prediction, while the lower dotted line indicates the two-dimensional classical bound.  The red (light gray) dot corresponds to the value of $I_{\rm DW}$ obtained experimentally in Ref.\ \cite{Polino2019}.}
\label{fig:IDW_results}
\end{figure}

Following Ref.\ \cite{Polino2019}, we computed $I_{\rm DW}$ for our model using the preparation settings $\phi_1 = 7\pi/4$, $\phi_2 = 5\pi/4$, $\phi_3 = \pi/2$ and measurement settings $\sigma_1 = \pi/2$, $\sigma_2 = 0$.  As before, we took $\sigma_0 = 1/\sqrt{2}$ and $\gamma = 1.95$ while varying $\alpha$.  The results are shown in Fig.\ \ref{fig:IDW_results}.  We find that the classical bound for two-dimensional models is surpassed for $|\alpha|^2 > 0.33$, with $|\alpha|^2 = 0.58$ giving the value $I_{\rm DW} = 3.82$ observed in Ref.\ \cite{Polino2019}.  For $|\alpha|^2 > 0.58$, we find that $I_{\rm DW}$ can exceed even the quantum bound.  Such exceedances, while often associated with post-quantum theories, are in fact a well known consequence of post-selection \cite{Marcovitch2007}.  Asymptotically, $I_{\rm DW}$ approaches the algebraic limit of five as $|\alpha|^2 \to \infty$.  The retrocausality bound $R_{\rm min}$ follows similar behavior.

It may seem curious that a manifestly causal model such as we have described would give a non-zero lower bound for the retrocausality measure.  Indeed, this would seem to contradict the assertion that $R_{Y \to \Lambda} \ge R_{\rm min}$.  However, the nonzero values of $R_{\rm min}$ are a direct consequence of $I_{\rm DW}$ exceeding the two-dimensional classical bound.  This, in turn, arises from the process of post-selection and gives rise to classical contextuality and, hence, the illusion of retrocausality.  Thus, there is no contraction between our results and the theoretical interpretation of $R_{\rm min}$.


\section{Preparation via Heralding}
\label{sec:RE}

The dimension witness has also been considered in a recent experiment by Huang \emph{et al.}\ using an entanglement source \cite{Huang2019}.  Instead of an attenuated laser, the experimenters used heralded detections on a parametric down conversion source of entangled light to produce the prepared state.  As in the Polino \emph{et al.}\ experiment of Ref.\ \cite{Polino2019}, polarization components are used as a surrogate for the two paths of an interferometer, and measurement proceeds in much the same way.

\begin{figure}[h]
\centering
\iffigures
\includegraphics[width=\columnwidth]{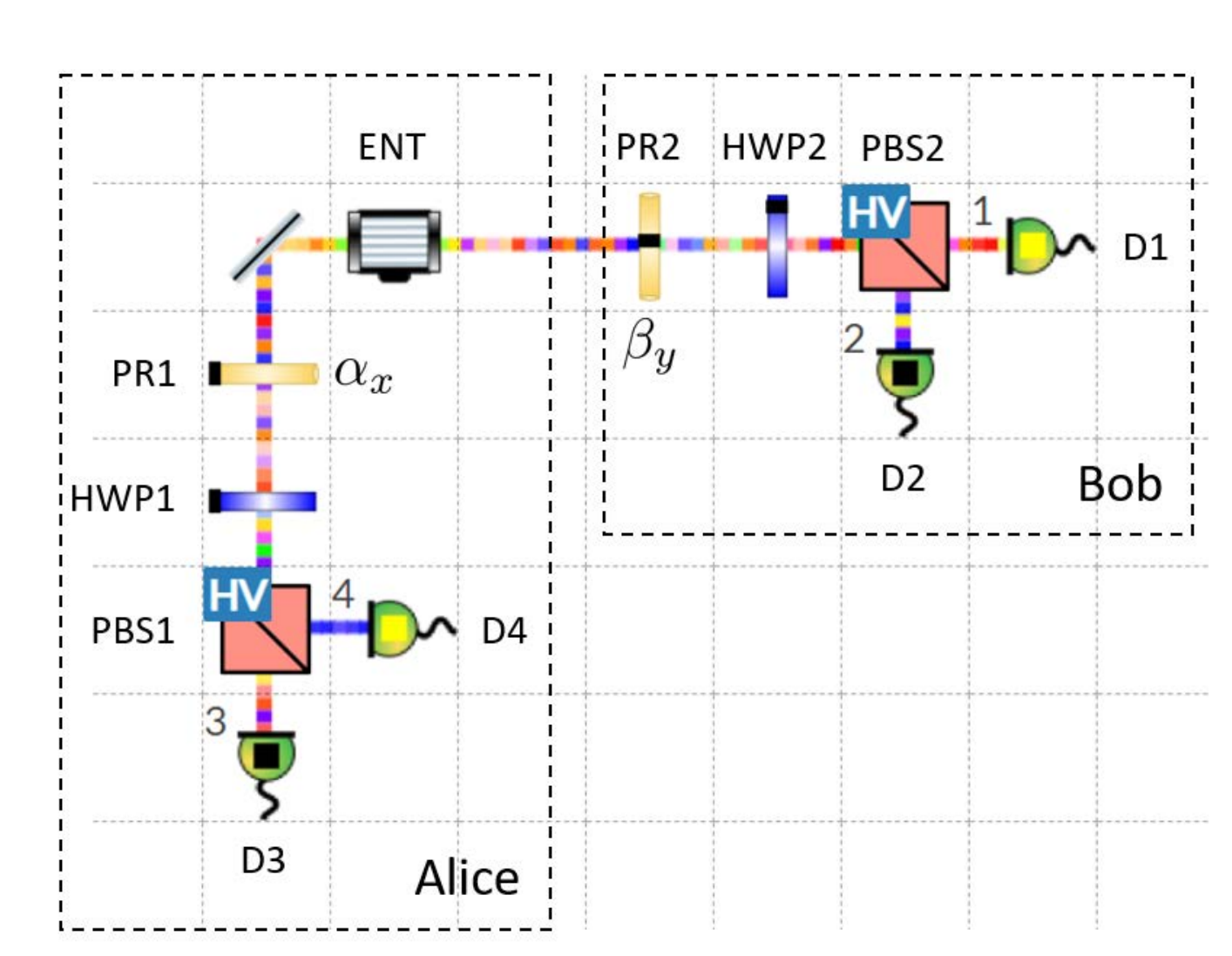}
\fi
\caption{(Color online) Experimental setup for an entanglement-based dimension witness measurement.}
\label{fig:EDW_setup}
\end{figure}

A simplified version of the experiment is shown in Fig.\ \ref{fig:EDW_setup}.  The entanglement source (ENT) is used to prepare the Bell state given by Eqn.\ (\ref{eqn:Bell}).  The setups for Alice and Bob are similar in that each uses a phase retarder (PR), a half-wave plate (HWP) set to $22.5^\circ$, a polarizing beam splitter (PBS) and a pair of detectors for each polarization mode.  Alice's phase retarder (PR1) is set to an angle $\alpha_x$, while Bob's (PR2) is set to $\beta_y$.  In the experiment of Ref.\ \cite{Huang2019}, $\alpha_1 = \pi/4, \alpha_2 = \pi/2$, while $\beta_1 = \pi/2, \beta_2 = 0$.  Also, the distance to Alice was made shorter than the distance to Bob, ensuring that setting changes by Alice cannot affect Bob.  Coincident detections are determined by the known relative delay.

The setup allows Alice to prepare four different states for Bob to measure, owing to the two values of $\alpha_x$ and the two detectors (D3 and D4) used by Alice for heralding.  Quantum mechanically, the state entering Alice's PBS is of the form $\ket{\Psi} = \ket{H} \otimes \ket{\psi_{3x}} + \ket{V} \otimes \ket{\psi_{4x}}$, where
\begin{align}
\ket{\psi_{3x}} &= \frac{\ket{H} + e^{i\alpha_x}\ket{V}}{\sqrt{2}} \\
\ket{\psi_{4x}} &= \frac{\ket{H} - e^{i\alpha_x}\ket{V}}{\sqrt{2}} \; .
\end{align}
Thus, a single detection on D3 may be construed as preparing the state $\ket{\psi_{3x}}$ for Bob, while one on D4 may be considered to prepare $\ket{\psi_{4x}}$.  Note that, since
\begin{equation}
\ket{\psi_{4x}} = \frac{\ket{H} + e^{i(\alpha_x+\pi)}\ket{V}}{\sqrt{2}} \; ,
\end{equation}
we may consider a herald on D4 as equivalent to Alice setting her phase retarder to the angle $\alpha_x +\pi$.

Following Bob's phase retarder and half-wave plate, the final state entering his PBS is
\begin{equation}
\ket{\psi'_{jxy}} = \frac{1}{2} \left( 1 + e^{i\phi_{jxy}} \right) \ket{H} + \frac{1}{2} \left( 1 - e^{i\phi_{jxy}} \right) \ket{V} \; ,
\end{equation}
where either $\phi_{3xy} = \alpha_x + \beta_y$ or $\phi_{4xy} = \alpha_x + \pi + \beta_y$ is used, depending upon whether Alice made a detection on D3 or D4, respectively.  The theoretical quantum prediction for the probability of a detection on, say, D1, given a single detection on either D3 or D4, is
\begin{equation}
q_{1j}(x,y) = \frac{1}{4} \left| 1 + e^{i\phi_{jxy}} \right|^2 \; .
\end{equation}
A tabulation of the different possible preparation and measurement bases is shown in Table \ref{tbl:bases}.

\begin{table}[h]
\begin{tabular}{cccc}
\hline
Basis & Herald & $(x,y)$ & $\phi_{jxy}$ \\ \hline
1 & D3 & (1,1) & $\pi/2+\pi/2 = \pi$ \\
2 & D3 & (2,1) & $\pi/4+\pi/2 = 3\pi/4$ \\
3 & D4 & (1,1) & $(\pi/2+\pi)+\pi/2 = 2\pi$ \\
4 & D4 & (1,1) & $(\pi/4+\pi)+\pi/2=7\pi/4$ \\
5 & D3 & (1,2) & $\pi/2$ \\
6 & D3 & (2,2) & $\pi/4$ \\
7 & D4 & (1,2) & $3\pi/2$ \\
8 & D4 & (1,2) & $5\pi/4$ \\ \hline
\end{tabular}
\caption{Enumeration of the eight preparation and measurement bases, as determined by Alice's heralding detector and the settings of the two phase retarders. }
\label{tbl:bases}
\end{table}

In our classical mode, the entanglement source is modeled by the random Jones vectors $\vec{a}$ and $\vec{b}$, where
\begin{equation}
\vec{a} = \begin{pmatrix} a_H \\ a_V \end{pmatrix}
= \sigma_0 \begin{pmatrix} z_{1H} \cosh r + z_{2H}^* \sinh r \\ z_{1V} \cosh r + z_{2V}^* \sinh r \end{pmatrix} \; \; \;
\end{equation}
and
\begin{equation}
\vec{b} = \begin{pmatrix} b_H \\ b_V \end{pmatrix}
= \sigma_0 \begin{pmatrix} z_{2H} \cosh r + z_{1H}^* \sinh r \\ z_{2V} \cosh r + z_{1V}^* \sinh r \end{pmatrix} \; .
\end{equation}
Local transformations by the phase retarders and half-wave plates yield
\begin{equation}
\vec{a} \xrightarrow{} \vec{a}'(x,y) = \frac{1}{\sqrt{2}} \begin{pmatrix} a_H + e^{i\alpha_x} a_V \\ a_H - e^{i\alpha_x} a_V \end{pmatrix} \; \; \;
\end{equation}
and
\begin{equation}
\vec{b} \xrightarrow{} \vec{b}'(x,y) = \frac{1}{\sqrt{2}} \begin{pmatrix} b_H + e^{i\beta_y} b_V \\ b_H - e^{i\beta_y} b_V \end{pmatrix} \; .
\end{equation}

Each of the polarizing beam splitters introduces an additional vacuum mode from the ZPF.  For Bob, we shall denote this is $\sigma_0(z_{3H}, z_{3V})^\mathsf{T}$, and for Alice we denote this $\sigma_0 (z_{4H}, z_{4V})^\mathsf{T}$.  The measurement events for a detection on each detector, regardless of the others, are therefore the following:
\begin{align}
D_1(x,y) &= \Bigl\{ |b'_H(x,y)| > \gamma \; \text{ or } \; \sigma_0 |z_{3V}| > \gamma  \Bigr\} \; , \\
D_2(x,y) &= \Bigl\{ |b'_V(x,y)| > \gamma \; \text{ or } \; \sigma_0 |z_{3H}| > \gamma  \Bigr\} \; , \\
D_3(x,y) &= \Bigl\{ |a'_H(x,y)| > \gamma \; \text{ or } \; \sigma_0 |z_{4V}| > \gamma  \Bigr\} \; , \\
D_4(x,y) &= \Bigl\{ |a'_V(x,y)| > \gamma \; \text{ or } \; \sigma_0 |z_{4H}| > \gamma  \Bigr\} \; .
\end{align}

We now consider the four coincident events $(D_1, D_3)$, $(D_1, D_4)$, $(D_2, D_3)$, $(D_2, D_4)$ for each of the possible values of $\alpha_x$ and $\beta_y$.  (For now, we drop the explicit dependence on $x$ and $y$.)  These events are given as follows:
\begin{align}
C_{13} &= D_1 \cap \bar{D}_2 \cap D_3 \cap \bar{D}_4 \\
C_{14} &= D_1 \cap \bar{D}_2 \cap \bar{D}_3 \cap D_4 \\
C_{23} &= \bar{D}_1 \cap D_2 \cap D_3 \cap \bar{D}_4 \\
C_{24} &= \bar{D}_1 \cap D_2 \cap \bar{D}_3 \cap D_4
\end{align}
The experiment considers coincident events
\begin{align}
C_1 &= C_{13} \cup C_{14} \\
C_2 &= C_{23} \cup C_{24}
\end{align}
such that $C_1$ is the event that Bob gets a click on just D1 and Alice gets a click on either D3 or D4 (but not both).

\begin{figure}[h]
\centering
\iffigures
\includegraphics[width=\columnwidth]{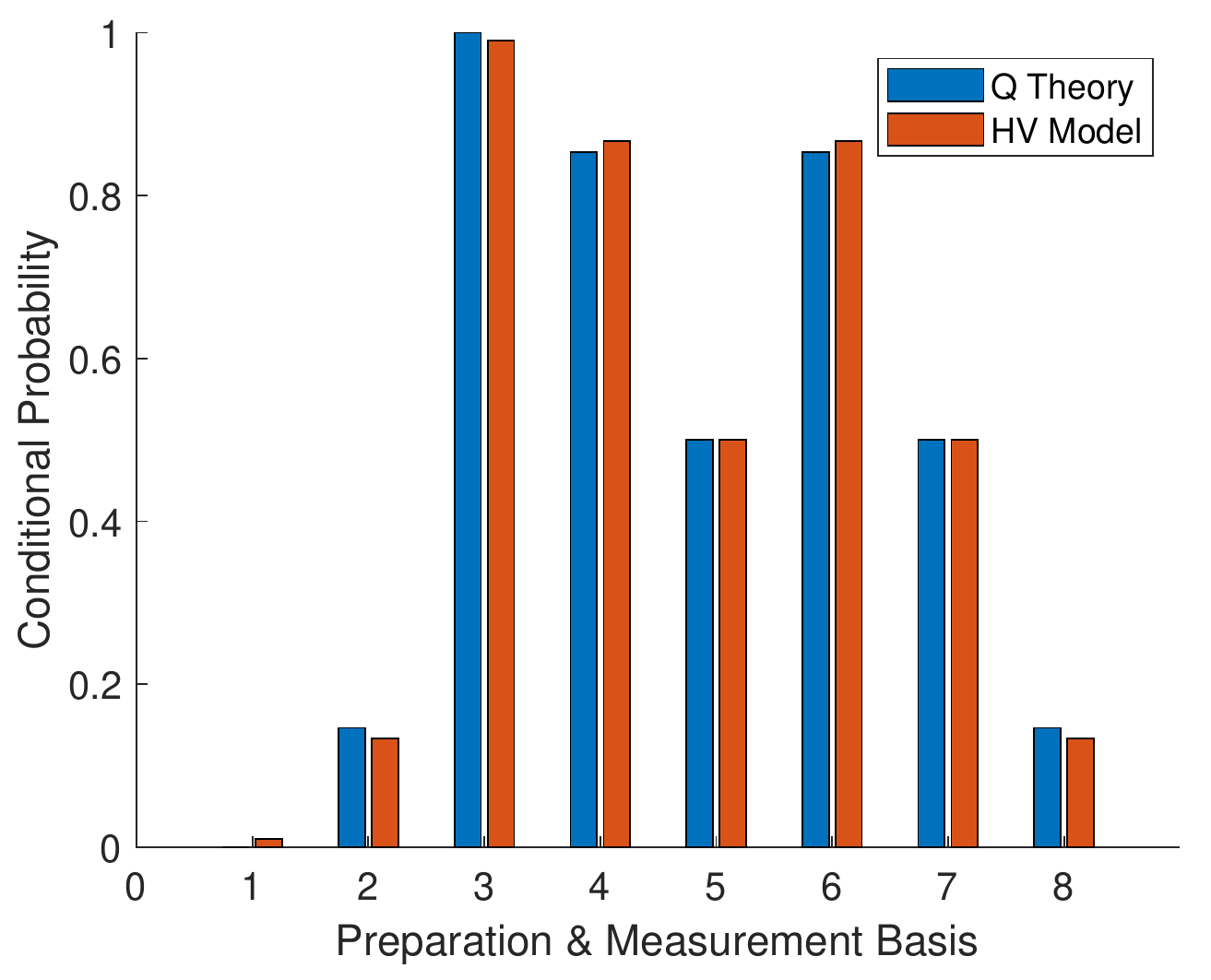}
\fi
\caption{(Color online) Plot of the conditional probability for each of the eight preparation and measurement bases, as defined in Table \ref{tbl:bases}.  The left blue (dark gray) bars are the theoretical quantum predictions $q_{1j}(x,y)$ , and the right red (light gray) bars are the values of $p_{1j}(x,y)$  determined numerically from our model.}
\label{fig:CondProb}
\end{figure}

From these events, we may define the conditional probabilities
\begin{equation}
p_{ij}(x,y) = \Pr[ C_{ij} | C_i ] = \frac{\Pr[C_{ij}(x,y)]}{\Pr[C_i(x,y)]} \; .
\end{equation}
Conditioning on coincident events is equivalent to adopting the fair-sampling assumption, which the experimenters in Ref.\ \cite{Huang2019} have used.  In Fig.\ \ref{fig:CondProb} we plot the numerically determined values of $p_{1j}(x,y)$, the conditional probabilities of detections on D1, and compare these against the theoretical quantum predictions $q_{1j}(x,y)$.  In the simulation, we used $r = 1$, $\gamma = 1.95$, $\sigma_0 = 1/\sqrt{2}$, and $N = 10^6$ random realizations.  We find that the agreement between theory and model predictions is comparable to or better than that found experimentally in Ref.\ \cite{Huang2019}.

\begin{figure}[h]
\centering
\iffigures
\includegraphics[width=\columnwidth]{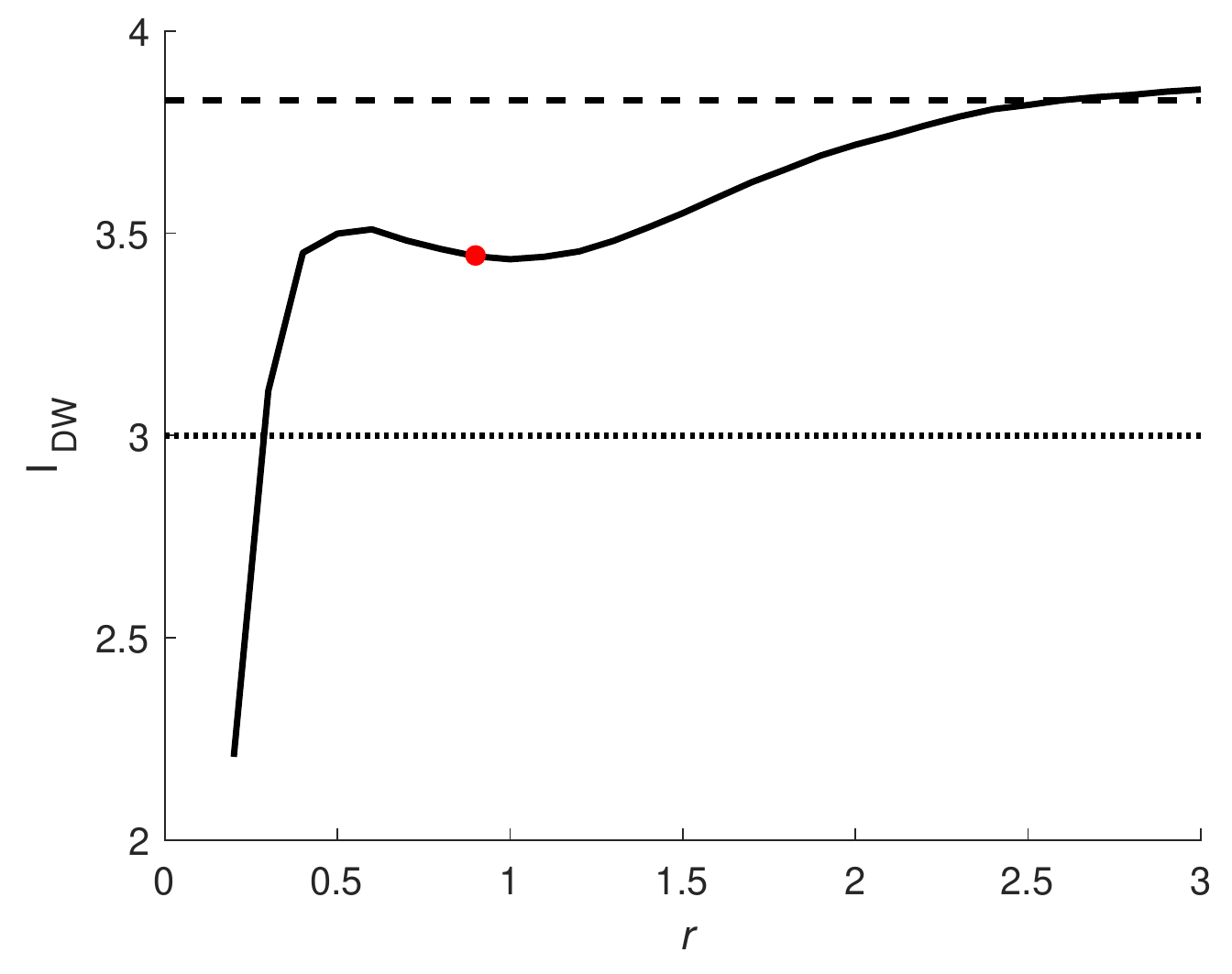}
\fi
\caption{(Color online) Plot of the dimension witness $I_{\rm DW}$ versus $r$.  The upper dashed line indicates the ideal quantum prediction, while the lower dotted line indicates the upper bound for a two-dimensional classical model.  The red (light gray) dot corresponds to the value of $I_{\rm DW}$ obtained experimentally in Ref.\ \cite{Huang2019}.}
\label{fig:EIDW_results}
\end{figure}

Finally, we compute the dimension witness
\begin{equation}
I_{\rm DW} = \Bigl| B_{311} + B_{312} + B_{321} - B_{322} - B_{411} \Bigr| \; ,
\end{equation}
where
\begin{equation}
B_{jxy} = p_{2j}(x,y) - p_{1j}(x,y) \; .
\end{equation}
Figure \ref{fig:EIDW_results} shows a plot of $I_{\rm DW}$ versus the squeezing parameter $r$.  (All other parameters are the same as those for Fig.\ \ref{fig:CondProb}.)  We see that for $r < 0.3$ the dimension witness falls below the threshold for a two-dimensional classical system.  At $r \approx 0.8$, $I_{\rm DW}$ achieves the value 3.445 observed in Ref.\ \cite{Huang2019}.  For $r > 2.6$, $I_{\rm DW}$ surpasses the theoretical quantum bound of $1+2\sqrt{2}$.  Larger values of $r$ result in numerical instability but suggest that $I_{\rm DW}$ continues to increase monotonically.

\section{Conclusion}
\label{sec:fin}

Current optical delayed-choice experiments, even those involving entangled light, can be understood from a strictly causal, classical perspective.  We illustrated this using a specific, physically motivated classical model with two key elements: (1) a reified zero-point field and (2) a deterministic threshold-based detector model.  This model is not restricted to delayed-choice experiments but is expected to be applicable to a wide range of quantum optical phenomena, although the precise domain of validity is not yet known.

The use of a dimension witness as a tool to distinguish classical from quantum systems was found to be inadequate due to its overly restrictive assumption of finite classical messaging.  The small class of hidden variable models that it is capable of ruling out is of no practical interest, as a simple classical device with analog messaging can easily spoof the witness.  Likewise the retrocausality measure, which is functionally related to the dimension witness, was found to provide no evidence for the presence of retrocausality or other nonclassical behavior, as nonzero values can easily be reproduced by a strictly causal classical model.

We found instead that the post-selection of desired measurement outcomes is critical to reproducing quantum behavior and appears to be what gives rise to the apparent causal (or retrocausal) behavior observed in delayed-choice experiments.  This is consistent with past experimental tests of quantum nonlocality that rely upon the fair-sampling assumption and thus may be susceptible to the detection loophole.  It may be possible that a delayed-choice experiment could be performed that avoids this detection loophole, but this has not yet been demonstrated.




\begin{acknowledgments}
This work was funded by the Office of Naval Research under Grant No.\ N00014-18-1-2107.
\end{acknowledgments}


\ifarxiv

\else
\bibliographystyle{apsrev4-1}
\bibliography{refs}
\fi

\end{document}